\input harvmac
\overfullrule=0pt
\parindent 25pt
\tolerance=10000
\input epsf

\newcount\figno
\figno=0
\def\fig#1#2#3{
\par\begingroup\parindent=0pt\leftskip=1cm\rightskip=1cm\parindent=0pt
\baselineskip=11pt
\global\advance\figno by 1
\midinsert
\epsfxsize=#3
\centerline{\epsfbox{#2}}
\vskip 12pt
{\bf Fig.\ \the\figno: } #1\par
\endinsert\endgroup\par
}
\def\figlabel#1{\xdef#1{\the\figno}}
\def\encadremath#1{\vbox{\hrule\hbox{\vrule\kern8pt\vbox{\kern8pt
\hbox{$\displaystyle #1$}\kern8pt}
\kern8pt\vrule}\hrule}}

\font\cmss=cmss10
\font\cmsss=cmss10 at 7pt
\def\rlx{\relax\leavevmode}
\def\inbar{\vrule height1.5ex width.4pt depth0pt}

  \def\frac#1#2{{#1\over #2}}

  \def\s{\sqrt}

  \def\p{\partial}

  \def\de{\partial}
  \def\db{\bar{\partial}}
  
  \def\lr{\leftrightarrow}
  \def\f {\frac}
  \def\ti{\tilde}
  \def\ap{\alpha}
  \def\pr{\propto}
  
  \def\ddd{\cdot\cdot\cdot}
  
  \def\la{\langle}
  \def\lb{\rangle}

  \def\vp{\varphi}

\def\p{\partial}

\def\CV{{\cal V}}
\def\CVp{{{\cal V}^\prime}}

\def\NN{{\cal N}}

\def\ie{{{\it i.e.}~}}
\def\IC{{\relax\,\hbox{$\inbar\kern-.3em{\rm C}$}}}
\def\IZ{\relax\ifmmode\mathchoice
{\hbox{\cmss Z\kern-.4em Z}}{\hbox{\cmss Z\kern-.4em Z}}
{\lower.9pt\hbox{\cmsss Z\kern-.4em Z}}
{\lower1.2pt\hbox{\cmsss Z\kern-.4em Z}}\else{\cmss Z\kern-.4em
Z}\fi}
\def\IR{\relax{\rm I\kern-.18em R}}
\def\fig#1#2#3{
\par\begingroup\parindent=0pt\leftskip=1cm\rightskip=1cm\parindent=0pt
\baselineskip=11pt
\global\advance\figno by 1
\midinsert
\epsfxsize=#3
\centerline{\epsfbox{#2}}
\vskip 12pt
\centerline{{\bf Figure \the\figno} #1}\par
\endinsert\endgroup\par}
\def\figlabel#1{\xdef#1{\the\figno}}

\def\pmb#1{\setbox0=\hbox{#1}%
\kern-.025em\copy0\kern-\wd0
\kern.05em\copy0\kern-\wd0
\kern-.025em\raise.0433em\box0 }
\font\cmss=cmss10
\font\cmsss=cmss10 at 7pt

\def\rlx{\relax\leavevmode}
\def\Cop{\relax\,\hbox{$\kern-.3em{\rm C}$}}
\def\Rop{\relax{\rm I\kern-.18em R}}
\def\Nop{\relax{\rm I\kern-.18em N}}
\def\Pop{\relax{\rm I\kern-.18em P}}

\def\Zop{\rlx\leavevmode\ifmmode\mathchoice{\hbox{\cmss Z\kern-.4em Z}}
{\hbox{\cmss Z\kern-.4em Z}}{\lower.9pt\hbox{\cmsss Z\kern-.36em Z}}
{\lower1.2pt\hbox{\cmsss Z\kern-.36em Z}}\else{\cmss Z\kern-.4em
Z}\fi}


\lref\Nstring{ M.~Ademollo {\it et al.}, ``Dual String With U(1)
Color Symmetry,'' Nucl.\ Phys.\ B {\bf 111}, 77 (1976);

For a review, refer to N.~Marcus, ``A Tour through N=2 strings,''
arXiv:hep-th/9211059.
}

\lref\OV{H.~Ooguri and C.~Vafa, ``Two-Dimensional Black Hole and
Singularities of CY Manifolds,'' Nucl.\ Phys.\ B {\bf 463}, 55
(1996) [arXiv:hep-th/9511164].
}

\lref\LechtenfeldIK{
  O.~Lechtenfeld and A.~D.~Popov,
  ``Closed N = 2 strings: Picture-changing, hidden symmetries and SDG
  hierarchy,''
  Int.\ J.\ Mod.\ Phys.\ A {\bf 15}, 4191 (2000)
  [arXiv:hep-th/9912154].
}

\lref\Hos{K.~Hosomichi, ``N = 2 Liouville theory with boundary,''
arXiv:hep-th/0408172.
}

\lref\OVN{H.~Ooguri and C.~Vafa, ``Selfduality And N=2 String
Magic,'' Mod.\ Phys.\ Lett.\ A {\bf 5}, 1389 (1990);
``Geometry of N=2 strings,'' Nucl.\ Phys.\ B {\bf 361}, 469 (1991).
}

\lref\BV{N.~Berkovits and C.~Vafa, ``N=4 topological strings,''
Nucl.\ Phys.\ B {\bf 433}, 123 (1995) [arXiv:hep-th/9407190].
}

\lref\OVA{
 H.~Ooguri and C.~Vafa,
  ``All loop N=2 string amplitudes,''
  Nucl.\ Phys.\ B {\bf 451}, 121 (1995)
  [arXiv:hep-th/9505183].
}

\lref\Ta{ T.~Takayanagi, ``Notes on S-matrix of non-critical N = 2
string,'' JHEP {\bf 0509}, 001 (2005) [arXiv:hep-th/0507065].
}

\lref\DHK{ J.~Distler, Z.~Hlousek and H.~Kawai, ``Superliouville
Theory As A Two-Dimensional, Superconformal Supergravity Theory,''
Int.\ J.\ Mod.\ Phys.\ A {\bf 5}, 391 (1990).
}

\lref\ABK{ I.~Antoniadis, C.~Bachas and C.~Kounnas, ``N=2
Superliouville And Noncritical Strings,'' Phys.\ Lett.\ B {\bf 242},
185 (1990).
}

\lref\TMV{T.~Takayanagi, ``$c < 1$ string from two dimensional black
holes,'' arXiv:hep-th/0503237.
}

\lref\KKL{E.~Kiritsis, C.~Kounnas and D.~Lust, ``A Large class of
new gravitational and axionic backgrounds for four-dimensional
superstrings,'' Int.\ J.\ Mod.\ Phys.\ A {\bf 9}, 1361 (1994)
[arXiv:hep-th/9308124].
}

\lref\tdsa{ D.~J.~Gross and N.~Miljkovic, ``A Nonperturbative
Solution Of D = 1 String Theory,'' Phys.\ Lett.\ B {\bf 238}, 217
(1990);
E.~Brezin, V.~A.~Kazakov and A.~B.~Zamolodchikov, ``Scaling
Violation In A Field Theory Of Closed Strings In One Physical
Dimension,'' Nucl.\ Phys.\ B {\bf 338}, 673 (1990);
P.~Ginsparg and J.~Zinn-Justin, ``2-D Gravity + 1-D Matter,'' Phys.\
Lett.\ B {\bf 240}, 333 (1990).
}

\lref\GLK{
 M.~Goulian and M.~Li,
 ``Correlation Functions In Liouville Theory,''
 Phys.\ Rev.\ Lett.\  {\bf 66}, 2051 (1991);
 Y.~Kitazawa,
 ``Gravitational descendents in Liouville theory,''
 Phys.\ Lett.\ B {\bf 265}, 262 (1991).
}

\lref\DK{P.~Di Francesco and D.~Kutasov, ``Correlation functions in
2-D string theory,'' Phys.\ Lett.\ B {\bf 261}, 385 (1991);
``World sheet and space-time physics in two-dimensional
(Super)string theory,'' Nucl.\ Phys.\ B {\bf 375}, 119 (1992)
[arXiv:hep-th/9109005].
}

\lref\MV{ J.~McGreevy and H.~Verlinde, ``Strings from tachyons: The
c = 1 matrix reloaded,'' JHEP {\bf 0312}, 054 (2003)
[arXiv:hep-th/0304224];
I.~R.~Klebanov, J.~Maldacena and N.~Seiberg, ``D-brane decay in
two-dimensional string theory,'' JHEP {\bf 0307}, 045 (2003)
[arXiv:hep-th/0305159];
J.~McGreevy, J.~Teschner and H.~L.~Verlinde, ``Classical and quantum
D-branes in 2D string theory,'' JHEP {\bf 0401}, 039 (2004)
[arXiv:hep-th/0305194];
A.~Sen, ``Open-closed duality: Lessons from matrix model,'' Mod.\
Phys.\ Lett.\ A {\bf 19}, 841 (2004) [arXiv:hep-th/0308068].
T.~Takayanagi and S.~Terashima, ``c = 1 matrix model from string
field theory,'' arXiv:hep-th/0503184.
}

\lref\type{T.~Takayanagi and N.~Toumbas, ``A matrix model dual of
type 0B string theory in two dimensions,'' JHEP {\bf 0307}, 064
(2003) [arXiv:hep-th/0307083];

M.~R.~Douglas, I.~R.~Klebanov, D.~Kutasov, J.~Maldacena, E.~Martinec
and N.~Seiberg, ``A new hat for the c = 1 matrix model,''
[arXiv:hep-th/0307195].
}

\lref\typemin{I.~R.~Klebanov, J.~Maldacena and N.~Seiberg, ``Unitary
and complex matrix models as 1-d type 0 strings,'' Commun.\ Math.\
Phys.\  {\bf 252}, 275 (2004) [arXiv:hep-th/0309168].
}

\lref\bosmin{ M.~R.~Douglas and S.~H.~Shenker, ``Strings In Less
Than One-Dimension,'' B {\bf 335}, 635 (1990);

D.~J.~Gross and A.~A.~Migdal, ``Nonperturbative Two-Dimensional
Quantum Gravity,'' Phys.\ Rev.\ Lett.\  {\bf 64}, 127 (1990);

E.~Brezin and V.~A.~Kazakov, ``Exactly Solvable Field Theories Of
Closed Strings,'' Phys.\ Lett.\ B {\bf 236}, 144 (1990).
}

\lref\ST{ A.~Strominger and T.~Takayanagi, ``Correlators in timelike
bulk Liouville theory,'' Adv.\ Theor.\ Math.\ Phys.\ {\bf 7}, 369
(2003) [arXiv:hep-th/0303221].
}

\lref\SC{V.~Schomerus, ``Rolling tachyons from Liouville theory,''
JHEP {\bf 0311}, 043 (2003) [arXiv:hep-th/0306026].
}

\lref\ESD{T.~Eguchi and Y.~Sugawara, ``Modular bootstrap for
boundary N = 2 Liouville theory,'' JHEP {\bf 0401}, 025 (2004)
[arXiv:hep-th/0311141];
}

\lref\ABKS{ O.~Aharony, M.~Berkooz, D.~Kutasov and N.~Seiberg,
``Linear dilatons, NS5-branes and holography,''
 {\bf 9810}, 004 (1998)
[arXiv:hep-th/9808149].
}

\lref\GK{A.~Giveon and D.~Kutasov, ``Comments on double scaled
little string theory,'' JHEP {\bf 0001}, 023 (2000)
[arXiv:hep-th/9911039].
}

\lref\HK{K.~Hori and A.~Kapustin, ``Duality of the fermionic 2d
black hole and N = 2 Liouville theory as mirror symmetry,'' JHEP
{\bf 0108}, 045 (2001) [arXiv:hep-th/0104202].
}

\lref\AFKS{O.~Aharony, B.~Fiol, D.~Kutasov and D.~A.~Sahakyan,
``Little string theory and heterotic/type II duality,'' Nucl.\
Phys.\ B {\bf 679}, 3 (2004) [arXiv:hep-th/0310197].
}

\lref\EY{T.~Eguchi and S.~K.~Yang, ``N=2 Superconformal Models As
Topological Field Theories,'' Mod.\ Phys.\ Lett.\ A {\bf 5}, 1693
(1990).
}

\lref\KPS{
  A.~Konechny, A.~Parnachev and D.~A.~Sahakyan,
  ``The ground ring of N = 2 minimal string theory,''
  Nucl.\ Phys.\ B {\bf 729}, 419 (2005)
  [arXiv:hep-th/0507002].
}

\lref\ADE{I.~K.~Kostov, ``Gauge invariant matrix model for the A-D-E
closed strings,'' Phys.\ Lett.\ B {\bf 297}, 74 (1992)
[arXiv:hep-th/9208053].
}

\lref\Ri{S.~Ribault, ``Knizhnik-Zamolodchikov equations and spectral
flow in AdS(3) string theory,'' JHEP {\bf 0509}, 045 (2005)
[arXiv:hep-th/0507114].
}

\lref\RT{ S.~Ribault and J.~Teschner, ``H(3)+ correlators from
Liouville theory,'' JHEP {\bf 0506}, 014 (2005)
[arXiv:hep-th/0502048].
}

\lref\GN{ G.~Giribet and Y.~Nakayama, ``The
Stoyanovsky-Ribault-Teschner map and string scattering amplitudes,''
arXiv:hep-th/0505203;
G.~Giribet, ``The String Theory on $AdS_3$ as a Marginal Deformation
of a Linear Dilaton Background,'' arXiv:hep-th/0511252.
}

\lref\AMT{S.~K.~Ashok, S.~Murthy and J.~Troost, ``Topological Cigar
and the c=1 String : Open and Closed,'' arXiv:hep-th/0511239.
}

\lref\Sto{A.V.Stoyanovsky, ``A relation between the Knizhnik--Zamolodchikov
and Belavin--Polyakov--Zamolodchikov systems of partial differential equations,''
arXiv:math-ph/0012013.
}

\lref\GOSZ{D.~Gluck, Y.~Oz and T.~Sakai, ``N = 2 strings on
orbifolds,'' JHEP {\bf 0508}, 008 (2005) [arXiv:hep-th/0503043].
}

\lref\ESP{T.~Eguchi and Y.~Sugawara, ``Modular invariance in
superstring on Calabi-Yau n-fold with A-D-E singularity,'' Nucl.\
Phys.\ B {\bf 577}, 3 (2000) [arXiv:hep-th/0002100].
}


\lref\KW{ I.~R.~Klebanov and R.~B.~Wilkinson, ``Critical Potentials
And Correlation Functions In The Minus Two-Dimensional Matrix
Model,'' Nucl.\ Phys.\ B {\bf 354}, 475 (1991).
}

\lref\DVVT{ R.~Dijkgraaf, H.~L.~Verlinde and E.~P.~Verlinde,
``Topological Strings In D < 1,'' Nucl.\ Phys.\ B {\bf 352}, 59
(1991).
}

\lref\WiT{ E.~Witten, ``On The Structure Of The Topological Phase Of
Two-Dimensional Gravity,'' Nucl.\ Phys.\ B {\bf 340}, 281 (1990).
}

\lref\WZW{ E.~Witten, ``The N matrix model and gauged WZW models,''
Nucl.\ Phys.\ B {\bf 371}, 191 (1992).
}

\lref\DW{ R.~Dijkgraaf and E.~Witten, ``Mean Field Theory,
Topological Field Theory, And Multimatrix Models,'' Nucl.\ Phys.\ B
{\bf 342}, 486 (1990).
}

\lref\RastelliPH{ L.~Rastelli and M.~Wijnholt, ``Minimal AdS(3),''
arXiv:hep-th/0507037.
}

\lref\DFMS{ L.~J.~Dixon, D.~Friedan, E.~J.~Martinec and
S.~H.~Shenker, ``The Conformal Field Theory Of Orbifolds,'' Nucl.\
Phys.\ B {\bf 282}, 13 (1987).
}

\lref\NN{ S.~Nakamura and V.~Niarchos, ``Notes on the S-matrix of
bosonic and topological non-critical strings,''
arXiv:hep-th/0507252.
}

\lref\KP{ I.~K.~Kostov, ``The ADE Face Models On A Fluctuating
Planar Lattice,'' Nucl.\ Phys.\ B {\bf 326}, 583 (1989);
I.~K.~Kostov, ``Strings with discrete target space,'' Nucl.\ Phys.\
B {\bf 376}, 539 (1992) [arXiv:hep-th/9112059];
I.~K.~Kostov and V.~B.~Petkova, ``Bulk correlation functions in 2D
quantum gravity,'' arXiv:hep-th/0505078.
}

\lref\Qiu{ Z.~a.~Qiu, ``Modular Invariant Partition Functions For
N=2 Superconformal Field Theories,'' B {\bf 198}, 497 (1987).
}

\lref\MKV{ S.~Mukhi and C.~Vafa, ``Two-dimensional black hole as a
topological coset model of c = 1 string theory,'' Nucl.\ Phys.\ B
{\bf 407}, 667 (1993) [arXiv:hep-th/9301083].
}

\lref\ET{T.~Eguchi and A.~Taormina, ``Unitary Representations Of N=4
Superconformal Algebra,'' Phys.\ Lett.\ B {\bf 196}, 75 (1987).
}

\lref\MO{J.~M.~Maldacena and H.~Ooguri, ``Strings in AdS(3) and
SL(2,R) WZW model. I,'' J.\ Math.\ Phys.\ {\bf 42}, 2929 (2001)
[arXiv:hep-th/0001053];
J.~M.~Maldacena, H.~Ooguri and J.~Son, ``Strings in AdS(3) and the
SL(2,R) WZW model. II: Euclidean black hole,'' J.\ Math.\ Phys.\
{\bf 42}, 2961 (2001) [arXiv:hep-th/0005183];
J.~M.~Maldacena and H.~Ooguri, ``Strings in AdS(3) and the SL(2,R)
WZW model. III: Correlation  functions,'' Phys.\ Rev.\ D {\bf 65},
106006 (2002) [arXiv:hep-th/0111180].
}

\lref\Zamo{A.~Zamolodchikov, ``On the three-point function in
minimal Liouville gravity,'' arXiv:hep-th/0505063.
}

\lref\Wa{ N.~P.~Warner, ``N=2 supersymmetric integrable models and
topological field theories,'' arXiv:hep-th/9301088.
}

\lref\RW{ L.~Rastelli and M.~Wijnholt, ``Minimal AdS(3),''
arXiv:hep-th/0507037.
}

\lref\GRT{D.~Gaiotto and L.~Rastelli, ``A paradigm of open/closed
duality: Liouville D-branes and the  Kontsevich model,'' JHEP {\bf
0507}, 053 (2005) [arXiv:hep-th/0312196];
}

\lref\BNW{M.~Bershadsky, W.~Lerche, D.~Nemeschansky and
N.~P.~Warner, ``Extended N=2 superconformal structure of gravity and
W gravity coupled to matter,'' Nucl.\ Phys.\ B {\bf 401}, 304 (1993)
[arXiv:hep-th/9211040].
}

\lref\KL{ K.~Li, ``Topological Gravity With Minimal Matter,'' Nucl.\
Phys.\ B {\bf 354}, 711 (1991);
``Recursion Relations In Topological Gravity With Minimal Matter,''
Nucl.\ Phys.\ B {\bf 354}, 725 (1991).
}

\lref\TopR{R.~Dijkgraaf, ``Intersection theory, integrable
hierarchies and topological field theory,'' arXiv:hep-th/9201003.
}

\lref\CHS{ C.~G.~.~Callan, J.~A.~Harvey and A.~Strominger,
``World sheet approach to heterotic instantons and solitons,''
Nucl.\ Phys.\ B {\bf 359}, 611 (1991).
}

\lref\Tesch{ J.~Teschner, ``On structure constants and fusion
rules in the SL(2,C)/SU(2) WZNW  model,''
Nucl.\ Phys.\ B {\bf 546}, 390 (1999)
[arXiv:hep-th/9712256].
}

\lref\minimal{ J.~Maldacena, G.~W.~Moore, N.~Seiberg and D.~Shih,
``Exact vs. semiclassical target space of the minimal string,'' JHEP
{\bf 0410}, 020 (2004) [arXiv:hep-th/0408039].
A.~Hashimoto, M.~x.~Huang, A.~Klemm and D.~Shih, ``Open / closed
string duality for topological gravity with matter,'' JHEP {\bf
0505}, 007 (2005) [arXiv:hep-th/0501141].
}

\lref\GRT{ D.~Gaiotto, L.~Rastelli and T.~Takayanagi, ``Minimal
superstrings and loop gas models,'' JHEP {\bf 0505}, 029 (2005)
[arXiv:hep-th/0410121];
A.~Kapustin, ``A remark on worldsheet fermions and double-scaled
matrix models,'' arXiv:hep-th/0410268.
}

\lref\AharonyKS{
 O.~Aharony,
 ``A brief review of 'little string theories',''
  Class.\ Quant.\ Grav.\  {\bf 17}, 929 (2000)
  [arXiv:hep-th/9911147].
}
\lref\KutasovUF{
  D.~Kutasov,
  ``Introduction to little string theory,''
{\it Prepared for ICTP Spring School on Superstrings and
Related Matters, Trieste, Italy, 2-10 Apr 2001}
}

\lref\GRK{
  D.~Gaiotto and L.~Rastelli,
  ``A paradigm of open/closed duality: Liouville D-branes and the  Kontsevich
  model,''
  JHEP {\bf 0507}, 053 (2005)
  [arXiv:hep-th/0312196].
}

\baselineskip 18pt plus 2pt minus 2pt

\Title{\vbox{\baselineskip12pt
\hbox{hep-th/0512112}\hbox{NSF-KITP-05-108}
   }}
{\vbox{\centerline{On the Connection between}
\medskip\centerline{$N=2$ Minimal String and $(1,n)$ Bosonic Minimal String}}}

\centerline{David A. Sahakyan\foot{E-mail:
sahakyan@physics.rutgers.edu} and Tadashi Takayanagi\foot{E-mail:
takayana@kitp.ucsb.edu}}

\bigskip
\bigskip

\centerline{${}^1$Department of Physics, Rutgers University,}
\centerline{ 126 Frelinghuysen rd., Piscataway NJ 08854,USA}
\medskip\centerline{ ${}^2$Kavli Institute for Theoretical
Physics}\centerline{University of California} \centerline{Santa
Barbara, CA 93106 USA}

\vskip .5in \centerline{\bf Abstract} We study the scattering
amplitudes in the $N=2$ minimal string or equivalently in the $N=4$
topological string on ALE spaces. We find an interesting connection
between the tree level amplitudes of the $N=2$ minimal string and
those  of the $(1,n)$ minimal bosonic string. In particular we show
that the four and five-point functions of the $N=2$ string can be
directly rewritten in terms of those of the latter theory. This
relation offers a map of physical states between these
two string theories. Finally we propose a possible matrix model dual
for the $N=2$ minimal string in the light of this connection.

\noblackbox

\Date{December, 2005}

\listtoc
\writetoc

\newsec{Introduction}

A particularly interesting set of string theories is obtained
by gauging $N=2$ superconformal symmetry on the worldsheet.
The resulting model is called the $N=2$ string \Nstring.
This string theory is very attractive since it has the
highest amount of supersymmetry on the world-sheet with the positive
critical dimension ($d=4$).
The simplest target space for the $N=2$ string is the four dimensional
flat spacetime with signature $(4,0)$ or $(2,2)$. This theory is referred as the
critical $N=2$ string. The theory with $(2,2)$ signature
can be solved exactly at least perturbatively.
The physical spectrum of it contains a single
massless scalar field whose dynamics is described by the mathematically 
beautiful theory of self-dual gravity \OVN.

Since the critical $N=2$ string has rich mathematical structure, it is
natural to hope that one can solve this string theory
even non-perturbatively. 
Indeed in a similar setups like two dimensional bosonic or $N=1$ type $0$
string theory we know
matrix model duals \refs{\tdsa, \type} and can solve the theory non-perturbatively.
These matrix models can be regarded as the open string theory of
D0-branes \MV\ via the open-closed duality. Hence it is very likely
that similar ideas can be applied to solve the $N=2$ string.

In this paper we study the $N=2$ string in a different background. 
The target space is the $N=2$ coset
space (Kazama-Suzuki model)
$SL(2,R)_{n}/U(1)\times SU(2)_{n}/U(1)$. In contrast to the critical $N=2$ string 
this theory has finite number of physical states.
We will refer to it
as the $N=2$ minimal string \refs{\KPS, \Ta}.
Again it is
natural to expect that this model has the matrix model dual
description as was true for the minimal bosonic \bosmin\ or type 0
string \typemin.

From higher dimensional string theory viewpoint, it is expected that
the BPS sector of the Little String Theory (LST)
(for recent review see \refs{\KutasovUF,\AharonyKS})
can be described by the $N=2$ minimal string \AFKS.
LST appears in various decoupling limits of string theories which contain  NS5
branes or singularities. This theory was extensively studied
using the holographically dual description \ABKS.
The simplest of LSTs are $5+1$ dimensional theories with sixteen supercharges.
They arise from the decoupling limit of $k$ type $IIA$ or type $IIB$
$NS5$-branes in flat space. The holographic description of these theories is given
by closed strings in the near horizon geometry of NS5 branes--the CHS background
\CHS.  Unfortunately, string theory in this background is strongly
coupled due to the presence of the linear dilaton.
One way to avoid this problem is to consider
the theory at a non-singular point in the moduli space. The simplest such
configuration corresponds to NS5 branes distributed on a circle. In this case
the CHS background gets deformed into \GK\
\eqn\CHSone{
\IR^{5,1}\times \left({SU(2)\over U(1)}\times {SL(2)\over U(1)}\right)/\IZ_k~,
}
where the $\IZ_k$ orbifolding ensures the R-charge integrality, \ie
imposes the GSO projection. We see that the non-trivial part of the background
\CHSone\ coincides with that of the minimal $N=2$ string.

It is instructive to consider the T-dual of the near horizon geometry of the NS5 branes.
Under T-duality the CHS background is mapped to a singular ALE\foot{The $N=2$ string on the smooth ALE spaces
defined by the orbifolds such as $\IC^2/\IZ_2$ was studied in \GOSZ.
See also the appendix A of the present paper for the analysis of
physical states for the $\IC^2/\IZ_N$ orbifold.} space 
(the \CHSone\ is mapped to the resolved ALE).
Then  the physical states of the $N=2$ string correspond to the deformation
preserving the hyper-K\"ahler structure of the ALE space.
In this sense the theory becomes topological and can be  equivalently described as
the $N=4$ topological string ($N=4$ TST) \BV.

We study the tree level scattering amplitudes in the $N=4$
topological string formulation using the equivalence between $N=2$
string and $N=4$ topological string theory \BV. We  find that
they are closely related to the amplitudes of the $(1,n)$ minimal
bosonic string\foot{This connection was also speculated in \Ta\ from
the analysis of three point functions in the $N=2$ string. We can
also find an earlier work \OV\ which implies a relation between the
$N=2$ string and $c<1$ bosonic string. It may also be closely
related to the recent observation made in \RW, where the equivalence
between the $N=2$ topological string on $AdS_3\times S^3$ and the
$(1,n)$ minimal string is argued from the cohomology analysis.} by
applying the recently found relation \refs{\Sto, \RT, \Ri, \GN}
between the correlation functions in
 the $SL(2,R)$ (or $H_3^+$) WZW model and those in the bosonic
Liouville theory. In particular we show that the four- and
five-point correlation functions of the $N=4$ topological string can be rewritten 
in terms of those of the $(1,n)$ minimal bosonic string. Moreover using this method
we are able also to match some classes of higher point correlators.
Even though our analysis is not exhaustive, it is
possible that these two string theories are actually equivalent or
at least closely related. In the end, based on this observation, we
 propose a candidate matrix model dual for the $N=2$
minimal string.

The paper is organized as follows. In section 2 we give a brief
review of the $N=2$ minimal string and analyze its physical states.
In section 3 we study the $N=4$ topological string on ALE spaces
which is equivalent to the $N=2$ minimal string. In section 4 we compute
the tree level scattering amplitudes and relate them to those of the
$(1,n)$ minimal string. In section 5 we discuss our results and propose 
a possible matrix model dual for $N=2$ minimal string.

\newsec{$N=2$ Minimal String}

\subsec{Notations}
In this subsection we set up notations. We are
interested in string theories with $SU(2)_{n}/U(1)\times
SL(2,R)_{n}/U(1)$ $N=2$ superconformal matter\foot{Here $n$ is the
level of the supercoset WZW model. The bosonic part of the coset is
given by $SU(2)_{n-2}/U(1)\times SL(2,R)_{n+2}/U(1)$.} (we will
often switch between $SL(2)/U(1)$ and the equivalent $N=2$ Liouville
descriptions).
The $N=2$ superconformal algebra reads
\eqn\ntwoalg{ \eqalign{&
[L_m,L_n]=(m-n)L_{m+n}+\f{c}{12}m(m^2-1)\delta_{m+n,0}\, ,\cr &
[L_{m},G^{\pm}_{r}]=\left(\frac{m}{2}-r\right)G^{\pm}_{m+r} \, ,\cr
& {[}L_{m}, J_{m}]=-nJ_{m+n} \, , \cr &
\{G^{+}_{r},G^{-}_{s}\}=2L_{r+s} + (r-s)J_{r+s} +
\frac{c}{3}\left(r^{2}-
 \frac{1}{4}\right)\delta_{r, -s} \, , \cr
& \{G^{+}_{r}, G^{+}_{s}\} = \{G^{-}_{r}, G^{-}_{s}\} = 0\, ,\cr &
{[}J_{n}, G_{r}^{\pm}]=\pm G^{\pm}_{r+n} \, ,\cr & {[}J_{m},
J_{n}]=\frac{c}{3}m\delta_{m, -n}\,, }} where $c$ is the central
charge, $L_n$ are the Virasoro algebra generators, $G^\pm_r$ are the
modes of the supercurrents ($r\in\IZ+1/2$ in the Neveu-Schwarz and
$r\in \IZ$ in the Ramond sectors) and $J_n$ are the modes of the
$U(1)$ R-current.

Let us start by setting notation for the $SU(2)/U(1)$ coset. The
central charge for this theory is \eqn\csu{ \hat c_{(su)}\equiv
{c\over 3}=1-{2\over n}~.} The NS sector $SU(2)/U(1)$ superconformal
primaries $\CV$ are labeled by $(j,m)$ quantum numbers, where
\eqn\jbound{\eqalign{ &2j\in \IZ,\,\, 0\leq 2j\leq n-2~,\cr &-j\leq
m\leq j,\,\,j+m\in \IZ~. }}

The dimension and the R-charge of the operator $\CV$ can be
expressed in terms of $m$ and $j$ as follows \eqn\dsursu{\eqalign{
\Delta_{(su)}&={j(j+1)\over n}-{m^2\over n}~,\cr R_{(su)}&=-{2m\over
n}~. }} In order to unify the description of NS and R-sectors it is
sometimes convenient to label operators in the $SU(2)/U(1)$ by three
quantum numbers $(j,m,s)$, where $s$ is the spectral flow parameter.
Not all $(j,m,s)$ are independent, there is a following equivalence
relation between them \eqn\eqivsu{ (j,m,s)\sim
({n-2\over 2}-j, m+{n-2\over 2}, s+2)~. } $s=0,2$ correspond to NS
operator, while $s=\pm 1$ to Ramond sector operators. The dimension
and the R-charge are \eqn\dsursus{ \eqalign{
\Delta_{(su)}&={s^2\over 8}+{j(j+1)\over n}-{(m+s/2)^2\over n}~,\cr
R_{(su)}&={s\over 2}-{2m+s\over n}~. }} In the NS-sector for $|m|<j$
only $s=0$ sector corresponds to superconformal primaries \eqn\sus{
\CV_{j,m}^{(s=\pm 2)}\sim G^{\pm}_{-1/2}\CV_{j,m\pm 1}^{(s=0)}\equiv
G^{\pm}_{-1/2}\CV_{j,m\pm1}~. } For $m=\pm j$, $\CV_{j,\pm
j}^{(s=\pm 2)}$ is actually primary. Indeed using \eqivsu\ we find
\eqn\chiralsu{ \CV_{j,\pm j}^{(s=\pm 2)}\sim \CV_{{n-2\over 2}-j,
\pm j\mp {n-2\over 2}}~. } The spectral flow $s\rightarrow s\pm 2$
can be realized by using the spectral flow operator \eqn\sfosu{
SFO_{(su)}^\pm=\CV_{{n-2\over 2},\mp {n-2\over 2}}~. } This can be
easily shown by using the $SU(2)$ fusion rules and the relation
\eqivsu.

We would also like to comment on the relation of the $N=2$ minimal
model to the $SU(2)$ WZW and the bosonic parafermions $SU(2)/U(1)$.
The $SU(2)_{n-2}$ primary $\Psi_{j,m}$ can be expressed in terms of
bosonic parafermion primary $V_{j,m}$ as follows\foot{ We normalized
the boson $Y_3$ such that $Y_3(z)Y_3(0)\sim -\log z$.} \eqn\parasu{
\Psi_{j,m}=V_{j,m} e^{i\sqrt{2\over n-2}m Y_3}~. } The $SU(2)$ currents
can be expressed using parafermionic currents $\psi_1$ and
$\psi_1^\dagger$ \eqn\sucurr{ \eqalign{ J^-&\sim\psi_1
\exp\left(-i\sqrt{2\over n-2}Y_3\right)~,\cr J^+&\sim\psi_1^\dagger
\exp\left( i\sqrt{2\over n-2}Y_3\right)~,\cr J^3&\sim i\s{\f{n-2}{2}}\
\p Y_3~. }} The spectral flowed primary operator is defined
\eqn\specsu{ \Psi_{j,m}^w=V_{j,m} e^{i\sqrt{2\over n-2}(m+{n-2\over
2}w)Y_3}~. } This has the eigenvalue $J_3=m+\f{n-2}{2}w$ and the
conformal dimension $\Delta=\f{j(j+1)}{n}+mw+\f{n-2}{4}w^2$. Note
that for the $SU(2)$, because of the Weyl identification $m\sim
m+(n-2)$ only spectral flow by $w=\pm 1$ are independent. Using the
identification of the quantum numbers
\eqn\jmw{ (j,m,w)\sim
({n-2\over 2}-j, {n-2\over 2}+m, w-1)~, } and the fact that the
operator $\Psi_{j,m}$ is $SU(2)$ current primary only for $-j\leq
m\leq j$ we find  that \eqn\currprim{ \Psi_{j, m}^{w=\pm 1}
={(J^\pm_{-1})}^{(j\pm m)}\Psi_{{n-2\over 2}-j, \pm{n-2\over 2}\mp
j}^{w=0}~. } One can rewrite the $N=2$ minimal model operators using
the bosonic parafermions (see \Qiu\ and the formula (4.47) and
(4.51) in \AFKS) \eqn\susypf{ \CV_{j,m}^s=V_{j,m}
e^{i{-2m+s{n-2\over 2}\over \sqrt{n(n-2)}}Y}\equiv V_{j,m}
e^{i\alpha_{m,s} Y}~. } Field $Y$ is essentially the bosonization of
the R-current. The $N=2$ supercurrents can also be rewritten\foot{
Again we normalized the boson $Y$ such that $Y(z)Y(0)\sim -\log z$.}
in term of $\psi_1$ and $\psi_1^\dagger$ \eqn\suprcurr{ \eqalign{
G^+&\sim\psi_1 \exp\left(i\sqrt{n\over n-2}Y\right)~,\cr
G^-&\sim\psi_1^\dagger \exp\left(-i\sqrt{n\over n-2}Y\right)~,\cr
J_R&\sim i\s{\f{n-2}{n}}\ \p Y~. }} Using these relations one can
explicitly check \sus. Now we are in position to express a general
correlator in the $N=2$ minimal model in terms of correlators of $SU(2)$
WZW 
\eqn\bossu{
\langle\CV_{j_1,m_1}^{s_1},\cdots,
\CV_{j_N,m_N}^{s_N}\rangle=\langle V_{j_1,m_1},\cdots
V_{j_N,m_N}\rangle \langle e^{i\alpha_{m_1,s_1}Y}\cdots
e^{i\alpha_{m_N,s_N}Y}\rangle~. }
Note that the R-charge conservation in the above formula requires
\eqn\rchrgsu{ \sum {m_i}-{n-2\over 4} s_i=0~, } which means that we
cannot lift the amplitude of bosonic parafermions to the amplitude
of $SU(2)$ primaries, unless $\sum s_i=0$. But it can be lifted into
an amplitude involving $\Psi_{j,m}^w$. Indeed if we have \eqn\wnds{
w_i=-s_i/2~, } the sum rule for $Y_3$ momentum will be satisfied and
we will get
\eqn\pftosu{ \langle\CV_{j_1,m_1}^{s_1},\cdots,
\CV_{j_N,m_N}^{s_N} \rangle={\langle\Psi_{j_1,m_1}^{w_1},\cdots,
\Psi_{j_N,m_N}^{w_N}\rangle \langle e^{i\alpha_{m_1,s_1}Y}\cdots
e^{i\alpha_{m_N,s_N}Y} \rangle\over\langle
e^{i\beta_{m_1,s_1}Y_3}\cdots e^{i\beta_{m_N,s_N}Y_3}\rangle}~, }
where (cr. \specsu) \eqn\bta{ \beta_{m,s}=\sqrt{2\over
n-2}(m-{n-2\over 4}s)~. } The case of chiral primary field ($m=-j$)
is a bit different since it can be represented as both
$s=0$ and $s=2$ operator \chiralsu.

It is straightforward to compute the free field parts in the
correlation function  \pftosu\ as follows \eqn\corfreeone{
\f{\langle \prod_{i=1}^N e^{i\alpha_{m_i,s_i}Y(z_i)}
\rangle}{\langle \prod_{i=1}^N
e^{i\beta_{m_i,s_i}Y_3(z_i)}\rangle}=\prod_{1\leq i<j\leq
N}\left(z_i-z_j\right)^{-\f{2}{n}(m_i-\f{n-2}{4}s_i)(m_j-\f{n-2}{4}s_j)}.}

\vskip.2in


The discussion for $SL(2,R)/U(1)$ case is essentially the same. Here
the superconformal operators $\CVp$ are labeled by quantum numbers
$(h,m,s)$, where $h$ is the $SL(2,R)$ spin, $m$ is $J_3$ quantum
number related to the R-charge and $s$ is the spectral flow
parameter. The formulae for the central charge,
the dimension and the R-charge of superconformal operators
can be obtained from the corresponding formulae for $SU(2)/U(1)$ by
taking $j\rightarrow -h$, $n\rightarrow -n$: 
\eqn\chat{ \eqalign{
&\hat c_{(sl)}=1+{2\over n}~,\cr &\Delta_{(sl)}={s^2\over
8}-{h(h-1)\over n}+{(m+s/2)^2\over n}~,\cr &R_{(sl)}={s\over
2}+{2m+s\over n}~. }} But there are also important differences. In
particular, the formulae \eqivsu, \sus, \chiralsu\ and \currprim,
which can be rewritten\foot{$\Phi^w_{h,m}$ denotes the spectral flowed
primary \MO\ of $SL(2,R)_{n+2}$ WZW model. It has the eigenvalue
$J_3=m-\f{n+2}{2}w$ and the conformal dimension
$\Delta=-\f{h(h-1)}{n}+mw-\f{n+2}{4}w^2$.} as \eqn\eqivsl{
\eqalign{& (h,m,s)\sim ({n+2\over 2}-h,m-{n+2\over 2}, s+2)~,\cr &
\CVp_{h,m}^{s=\pm 2}=G^{\pm}_{-1/2} \CVp_{h,m\pm 1}^{s=0}~, \cr &
\CVp_{h,\mp h}^{s=\pm 2}\sim \CVp_{\f{n+2}{2}-h~,\mp
h\pm\f{n+2}{2}}^{s=0}~, \cr & \Phi^{w=\pm
1}_{h,m}=(J^{\pm}_{-1})^{m-h}\
\Phi^{w=0}_{\f{n+2}{2}-h,\mp\f{n+2}{2}\pm h}~. }} only apply
to the operators corresponding to the discrete ($h\in \IR$, $h-m\in
\IZ$ or $h+m\in \IZ$) (chiral or anti-chiral) and degenerate
representations ($2h\in -\IZ_{>0}$, $-|h|\leq m\leq |h|$) of $SL(2)$.
The spectral flow operator in the $SL(2)/U(1)$ is \eqn\spectot{
SFO_{(sl)}^\pm=\CVp_{{n+2\over 2},\pm {n+2\over 2}}~. } We would
like to derive the analog of the formula\foot{In the same way as
before we normalized the free boson $X_3$ and $X$ such that
$X(z)X(0)\sim X_3(z)X_3(0)\sim -\log z$.} \pftosu\ for $SL(2)/U(1)$
\eqn\pftosl{ \langle\CVp_{h_1,m_1}^{s_1},\cdots,
\CVp_{h_N,m_N}^{s_N} \rangle={\langle\Phi_{h_1,m_1}^{w_1},\cdots,
\Phi_{h_N,m_N}^{w_N}\rangle \langle e^{i\alpha'_{m_1,s_1}X}\cdots
e^{i\alpha'_{m_N,s_N}X}\rangle\over\langle
e^{\beta'_{m_1,s_1}X_3}\cdots e^{\beta'_{m_N,s_N}X_3}\rangle}~, }
where \eqn\betsl{\eqalign{ &\beta'_{m,s}=\sqrt{2\over
n+2}(m+{n+2\over 4}s)~,\cr &\alpha'_{m,s}=-{2m+s{n+2\over 2}\over
\sqrt{n(n+2)}}~,\cr &w=-{s\over 2}~. }} Finally we can also find the
following formula analogous to \corfreeone\ \eqn\corfreetwo{
\f{\langle \prod_{i=1}^N e^{i\alpha'_{m_i,s_i}X(z_i)}
\rangle}{\langle \prod_{i=1}^N
e^{i\beta'_{m_i,s_i}X_3(z_i)}\rangle}=\prod_{1\leq i<j\leq
N}\left(z_i-z_j\right)^{\f{2}{n}(m_i+\f{n+2}{4}s_i)(m_j+\f{n+2}{4}s_j)}~.}

\vskip.3in

\subsec{$N=2$ Minimal String Theory}

The $N=2$ minimal string theory is defined by coupling the $N=2$
minimal matter to the $N=2$ Liouville theory \refs{\DHK, \ABK}, and
gauging the world-sheet $N=2$ superconformal symmetry. The $N=2$
minimal model is equivalent to the $N=2$ coset $SU(2)/U(1)$, while
the $N=2$ Liouville theory is T-dual to the $N=2$ coset
$SL(2,R)/U(1)$ via the supersymmetric FZZ duality  \refs{\GK,\HK}.
Therefore the target space of the $N=2$ minimal string theory is
given by the following product of two $N=2$ superconformal cosets
\refs{\KKL,\OV} (as we will see below this product actually enjoys
enhanced $N=4$ superconformal symmetry on the world-sheet)
\eqn\cosetdefg{ \left[\f{SL(2,R)_{n}}{U(1)}\times
\f{SU(2)_{n}}{U(1)}\right]/\IZ_n~,} with the total central charge
\eqn\cmat{ \hat c_{tot}=2~,} where the $\IZ_n$ orbifold in
\cosetdefg\ ensures integral R-charges of states as required by the
modular invariance of the $N=2$ string. The $N=2$ string contains
the usual $(b,c)$ conformal ghosts, two pairs of superconformal
ghosts $(\beta^\pm,\gamma^\mp)$ and an additional $(\tilde b, \tilde
c)$ fermionic ghost system, which arises from gauging the R-current.
One can check that the total central charge of the $N=2$ ghost
system is \eqn\centgh{ \hat c_{gh}=-2~, } which precisely matches
the the central charge of the matter sector. It is convenient to
bosonize the ${\beta, \gamma}$ system as follows \eqn\bgbos{
   \beta^\pm\sim e^{-\phi_\mp} \p\xi^\pm;\qquad
   \gamma^\pm\sim\eta^\pm e^{\phi_\pm}~.
}
The physical states of the $N=2$ string theory are elements of cohomology groups
of the BRST operator
\eqn\brsop{
Q_{BRST}={1\over 2\pi i} \oint dz\, j_{BRST}\,,
}
where the BRST current takes the form
\eqn\BRSTcur{
\eqalign{
j_{BRST} =& cT +  \eta_{-}e^{\phi_{-}}G^{+} +
\eta_{+}e^{\phi_{+}}G^{-}
 + \tilde cJ^{m} + \cr
& {1\over 2}[cT^{gh} + \eta_{-}e^{\phi_{-}}G^{+}_{gh}
+ \eta_{+}e^{\phi_{+}}G^{-}_{gh} +  \tilde cJ^{gh}]\,.
}}
The BRST current has non-singular OPE with the two picture number currents
\eqn\picnum{
  j_{\pi+}=-\eta^+\xi^--\p\phi_+;\qquad j_{\pi-}=-\eta^-\xi^+-\p\phi_- \, ,
}
and the ghost number current
\eqn\ghnum{
j_{gh}=-bc-\ti b\ti c+\eta^+\xi^-+\eta^-\xi^+~.
}
Hence the corresponding cohomology groups are labeled
by the ghost number and the picture
numbers $(\Pi_{+}, \Pi_{-})$.
One can  define two picture raising operators
 \eqn\pco{
PCO^{\pm}=\{Q,\xi^\pm\}= c\p\xi^\pm+e^{\phi^{\mp}}(G^{\pm}-2\eta^\pm
e^{\phi^\pm}b\pm 2\p(\eta^\pm e^{\phi^\pm}) \tilde b\pm\eta^\pm
e^{\phi^\pm}\p\tilde b ) \, . } It is known \LechtenfeldIK\ that
unlike in the $N=1$ superstrings, the picture raising operators
\pco\ are not isomorphisms of the cohomology groups at different
pictures, which in general complicates analysis of cohomologies.

Let us now discuss the cohomologies of the BRST operator at lower
pictures and ghost number one. We will follow closely the discussion in \KPS.
The BRST invariant operator in the
standard $(-1,-1)$ picture can be written as \foot{In this section
we write only the chiral part of the operators to keep the
expressions simple.} \eqn\vjm{ V_{j,m}^{(-1,-1)}=c
\CV_{j,m}\CVp_{h=-j,m} e^{-\phi_+}e^{-\phi_-}. } The operator ${\cal
V}$ and ${\cal V}^\prime$ should be primaries of the corresponding
${\cal N}=2$ algebras, in order for $\cal O$ to be BRST closed. This
operator, for generic $m$, has images in the lower pictures. This
can be shown using \pco. The result is \eqn\pr{ V_{j,m}^{(0,-1)}=c
G^{-}_{-{1\over 2}}(\CV_{j,m}\CVp_{-j,m})e^{-\phi_-}. }
In general the BRST cohomologies in this picture have the following
form 
\eqn\tldc{ \tilde{\cal O}=ce^{-\phi_-} {\tilde{\cal V}}{\tilde{\cal V}}^\prime, } 
where $ {\tilde{\cal V}}{\tilde{\cal
V}}^\prime$ satisfies \eqn\physst{ \eqalign{ &G^{+}_r( {\tilde{\cal
V}}{\tilde{\cal V}}^\prime)=0~,\cr &G^{-}_{r-1}( {\tilde{\cal
V}}{\tilde{\cal V}}^\prime)=0,\quad r>0~,\cr &\Delta( {\tilde{\cal
V}}{\tilde{\cal V}}^\prime)={1\over 2}~,\cr &q( {\tilde{\cal
V}}{\tilde{\cal V}}^\prime)=-1~.}} These conditions are indeed
satisfied for the $V_{j,m}^{(0,-1)}$, but it is easy to see that in
this picture there are additional BRST invariant operators
corresponding to the fields which are anti-chiral in $SU(2)/U(1)$
and $SL(2,R)/U(1)$ separately \eqn\acc{ \phi_h^{(0,-1)}=ce^{-\phi_-}
\CV_{{n\over 2}-h,{n\over 2}-h} \CVp_{h,-h}~. } It is clear that this
operator is not of the form \pr, hence does not have preimage in the
$(-1,-1)$ picture. In this paper we will be mainly interested in the
physical operators of this type. Vertex operators based on chiral
fields in the $SU(2)/U(1)$ and $SL(2)/U(1)$ can be constructed in
the similar manner. These vertex operators live naturally in the
$(-1,0)$ picture \eqn\cc{ \phi_h^{(-1,0)}=ce^{-\phi_+} \CV_{{n\over
2}-h,-{n\over 2}+h} \CVp_{h,h}~. } The operators in the $(0,-1)$ and
$(-1,0)$ pictures are related via spectral flow, which can be
realized using the following BRST invariant operators \eqn\sfobrst{
(S^\pm)^2=e^{\pm\ti b c}
e^{\phi_\pm}e^{-\phi_\mp}SFO^\mp_{(su)}SFO^\mp_{(sl)}~. } Indeed
using \sfobrst\ one can immediately see that \eqn\chantichrel{
\phi_{{n+2\over 2}-h}^{(0,-1)}=(S^+)^2\phi_h^{(-1,0)}~. } In order
to compute correlation functions one also needs the integrated form
of these operators in $(-1,0)$ and $(0,0)$ picture 
\eqn\intgroo{
\eqalign{ &\phi_{h,int}^{(0,-1)}=\int e^{-\phi_-} \CV_{{n\over
2}-h,{k\over 2}-h} \CVp_{h,-h}~,\cr &\phi_{h,int}^{(-1,0)}=\int
e^{-\phi_+} \CV_{{n\over 2}-h,-{n\over 2}+h} \CVp_{h,h}~. }} 
Applying
\pco\ one can easily find the form of these operators in the $(0,0)$
picture \eqn\intgroopico{ \eqalign{ PCO^+\phi_{h,int}^{(0,-1)}=\int
c\p\xi^+e^{-\phi_-} \CV_{{n\over 2}-h,{n\over 2}-h} \CVp_{h,-h}+
\int G^+_{-1/2}(\CV_{{n\over 2}-h,{n\over 2}-h} \CVp_{h,-h})~,\cr
PCO^-\phi_{h,int}^{(-1,0)}=\int c\p\xi^-e^{-\phi_+} \CV_{{n\over
2}-h,-{n\over 2}+h} \CVp_{h,h}+ \int G^-_{-1/2}(\CV_{{n\over
2}-h,-{n\over 2}+h} \CVp_{h,h})~. }} As we will see below it is
sufficient for our purposes to retain only the second term in these
expressions. Restoring the $\bar z$ dependence we find
\eqn\ingrfull{ \eqalign{ &PCO^+\bar {PCO^+}\phi_{int}^{(0,-1)}=\int
d^2 z G^+_{-1/2}\bar G^+_{-1/2}(\CV_{{n\over 2}-h,{n\over 2}-h}
\CVp_{h,-h})+\cdots~,\cr &PCO^-\bar {PCO^-}\phi_{int}^{(-1,0)}=\int
d^2 z G^-_{-1/2}\bar G^-_{-1/2}(\CV_{{n\over 2}-h,-{n\over 2}+h}
\CVp_{h,h})+\cdots~. }}


It is also interesting to note that the operators  $PCO^\pm$ may map
the BRST non-trivial operators into the BRST trivial ones. This
statement in the free $N=2$ string translates into the statement
that picture changing acts as an isomorphism of the BRST cohomology
groups only for the states with non-zero momentum. The simplest
example of such phenomenon is the operator \eqn\calo{ {\cal
O}=ce^{-\phi_-}e^{-\phi_+}{\bf 1}~. } This operator is mapped into
zero by the action of both $PCO^\pm$. In general BRST non-trivial
operators of the type \eqn\calone{ {\cal
O}=ce^{-\phi_-}e^{-\phi_+}{\cal V}{\cal V}^\prime~, } where $\cal V$
and ${\cal V}^\prime$ are (anti-)chiral primaries of $SU(2)/U(1)$
and $SL(2)/U(1)$ respectively are mapped into zero by the action of
$PCO^+$ ($PCO^-$).

Finally, to check the consistency of the string theory we need to
find a modular invariant partition function. We worked this out in
the appendix C.

\newsec{Equivalent Description as $N=4$ Topological String on ALE
Spaces}

\subsec{Exhibiting the $N=4$ structure of $SU(2)/U(1)\times
SL(2)/U(1)$ $N=2$ theory}

As noted in \refs{\ET,\BV} any $N=2$ superconformal theory with the
central charge \eqn\crit{ \hat c=2~, } automatically has $N=4$
superconformal symmetry. The $N=4$ superconformal algebra is defined
by the energy-momentum tensor $T$, the four supercurrents
$(G^+,\ti{G}^+,G^-,\ti{G}^-)$ and the $SU(2)$ currents
$(J^{++},J^{--},J^3)$. $J_3$ is identified with the R-current in the
$N=2$ subalgebra. $(G^+,\ti{G}^+)$ have $J_3$ charge $+1$ while
$(G^-,\ti{G}^-)$ have the $-1$ charge.

For the particular value \crit\ of the central charge the
spectral flow operators $SFO^\pm$ have the R-charge $\pm 2$ and
dimension one and hence can serve\foot{We work in the notations of
\BV.} as $J^{++}$ and $J^{--}$ currents of the $N=4$. In the case at
hand \eqn\jpp{ J^{\pm\pm}=SFO^\pm\equiv SFO_{(su)}^\pm
SFO_{(sl)}^\pm~. } Also $\tilde G^+$ and $\ti{G}^-$ can be found as
\eqn\tldg{ \tilde G^+_r=[J_0^{++},G^-_r]~, \ \ \
\ti{G}^-_r=[J_0^{--},G^+_r].}

 It is well
known that the deformations of the theory which respect the $N=2$
structure are in one to one correspondence with chiral operators of
R-charge $1$ and dimension $1/2$. A deformation by an operator $V$
respects the $N=4$ structure\foot{The third and fourth conditions
are always true in the unitary SCFT when $V$ is a chiral field with
the R-charge $+1$. Here we are not assuming the unitarity of the
underlining SCFT.} iff \BV \eqn\resp{ G^+_{-1/2}V=0,\,\, \tilde
G^+_{-1/2}V=0, \,\, J^{++}_0V=0, \,\, J^{--}_0V{\rm\,\, is\,\,
anti-chiral}~. } It is instructive to check that the deformations of
the type \eqn\phij{ V_h=\CV_{{n\over 2}-h,-{n\over 2}+h} \CVp_{h,h}~,
} respect the $N=4$ structure. The first of the \resp\ relations is
obviously satisfied, since both $SU(2)/U(1)$ and $SL(2)/U(1)$
operators are chiral primaries. Using \eqivsu\ one can see also
that\foot{We define the complex conjugate $\bar{V}_h$ of $V_h$ by
$\bar{V}_h=\CV_{{n\over 2}-h,{n\over 2}-h} \CVp_{h,-h}$, flipping
the sign of $J_3$.}
 \eqn\jmm{ J^{--}_0 V_h=\bar V_{{n+2\over 2}-h}~,
} hence the fourth relation is also correct. It is not hard to show
that the second and third relations also hold. We notice that the
$V_h$ deformations are in one to one correspondence with the
accidental cohomologies $\phi_h^{(-1,0)}$ of the $N=2$ string \acc.
One can similarly check that the deformations which correspond to
the $V_{j,m}^{(-1,0)}$ (see \pr) \eqn\gp{
G^+_{-1/2}(\CV_{j,m},\CVp_{h=-j,m})\equiv G^+_{-1/2}V_{j,m}~, } also
respect the $N=4$ structure. 

\subsec{Geometrical Interpretation}

The target space defined by the $\hat{c}=2$ coset space \cosetdefg\
can also be regarded as the resolved $A_{n-1}$ singularity \refs{\OV, \GK}\foot{ To
make the connection with the coset CFT \cosetdefg\ clearer we can
express it in the homogeneous coordinates as $x^n+y^2+z^2-\mu_{sl}\
u^{-n}=0$. The $N=2$ Landau-Ginzburg model for $W=x^n$ and
$W=u^{-n}$ is equivalent to the $SU$ and $SL$ part of the coset
\OV.} \eqn\algc{ x^n+y^2+z^2=\mu_{sl}~.}  The deformation
parameter $\mu_{sl}$ is equal to the $N=2$ cosmological constant in
the dual $N=2$ Liouville theory. This background can be regarded as 
regularized CHS geometry \refs{\CHS, \KKL}. Indeed it is
well-known that it is T-dual to the near horizon geometry of $n$
NS5-branes situated on a circle of radius $r_0\sim
(\mu_{sl})^{\f{1}{n}}$ \GK.

The K\"ahler and complex structure deformations of the ALE space are described
by the $(c,c)$, $(a,a)$, $(a,c)$ and $(c,a)$ rings of the SCFT \cosetdefg.
There are $4(n-1)$ such deformations for the $A_{n-1}$ ALE space--two complex structure
and two K\"ahler deformations for each $2$-cycle.


 
These deformations are described in the  SCFT by the
interaction terms on the world-sheet (we show only left-moving
part)\eqn\defint{\sum_{i=1}^{n-1}\left(t^{(i)}_{1L}\int
G^-_{-1/2}V^{(i)}+t^{(i)}_{2L}\int \ti{G}^-_{-1/2}V^{(i)}\right)~,}
where the second term can be written as 
$G^+\bar{V}$\foot{ Indeed using \jmm\ and \tldg\ one can show that
$G^{+}\bar{V}_{h}=\ti{G}^-V_{\f{n+2}{2}-h}$.}. 
By adding the right-moving sector, we have four parameters\foot{In
the notation in \refs{\BV, \OVA} these correspond to the twistor
variables $u_{1L}, u_{2L},u_{1R}, u_{2R}$.} $t^{(i)}_{1L},
t^{(i)}_{2L},t^{(i)}_{1R}$ and $t^{(i)}_{2R}$ for each
$i=1,2,...,n-1$. The four combinations $t^{(i)}_{1L}t^{(i)}_{1R},
t^{(i)}_{1L}t^{(i)}_{2R}, t^{(i)}_{2L}t^{(i)}_{1R}$ and
$t^{(i)}_{2L}t^{(i)}_{2R}$ are the moduli corresponding to the
$i-$th 2-cycle. 

As we have seen in the previous subsection there are  $n-1$ operators 
(again we show only the left moving part)
of the type \phij\ in our model \cosetdefg 
\eqn\deformphys{V_{h}\ \ \ (2h-1=1,2,...,n-1)~.}
After including the right movers we get exactly $4(n-1)$ operators. 
These operators are 
responsible for the  K\"ahler and complex structure deformations of the ALE space.
Notice that this
finite number of allowed states come from the familiar bound
\jbound\ of $j$ in the $SU(2)_{n-2}$ model and it is also consistent
with the unitarity bound of $SL(2,R)_{n+2}$ model \MO. Similarly, we
can check that $n-1$ twisted sector states exist in the orbifold
$\IC^2/\IZ_n$ which satisfy \resp\ as we show in the appendix A
(see also \GOSZ).

In addition to the states \deformphys\ there exist other
deformations \gp\ which come from $(-1,-1)$ picture states. At
present we do not have any clear geometrical interpretation of them.
As we cannot find any such states as \gp\ in the orbifold case
$\IC^2/\IZ_n$, the number of this type of states may not be
conserved in the geometrical deformations of the theory. Thus we
will concentrate on the states \deformphys\ in the rest of this
paper.


\subsec{Definition of $N=4$ Topological String}

Sometimes it is useful to employ the $N=4$ topological string description
which is known to be equivalent to the $N=2$ string \BV.

The $N=4$ topological string is defined as follows. Consider a
$\hat{c}=2$ $N=4$ SCFT and perform the usual topological twist $T\to
T+\f{1}{2}\de J^3$ \refs{\EY,\BV}. After the twist the operators
$G^+$ and $\ti{G}^+$ have the conformal dimension $\Delta=1$, while
$G^-$ and $\ti{G}^-$ have $\Delta=2$. Since the former ones satisfy
$(G^+_0)^2=(\ti{G}^+_0)^2=\{G^+_0,\ti{G}^+_0\}=0$, they
  behave like BRST operators.

The physical states have R-charge one and are the top components of
an $SU(2)$ doublet, whose bottom components are anti-chiral with
$R=-1$. They satisfy
 \eqn\physco{G^+_{0}V=\ti{G}^+_{0}V=0~,}
and are subject to equivalence relation $V\sim
V+G^+_{0}\ti{G}^+_{0}\chi$.
Conditions \physco\ are equivalent to \resp\ before we take the topological
twist. Hence the physical states of the $N=4$ string are in one to one correspondence
with the $(-1,0)$ picture states of the $N=2$ string.


\subsec{Physical States in $N=4$ Topological String}

Here we summarize physical states in $N=4$ topological string on the
ALE spaces. We present the vertex operators from the NS-sector
viewpoint before the topological twisting.

In this paper we are mainly interested in the operators of the type
\phij, which are separately chiral in the $SU(2)/U(1)$ and
$SL(2)/U(1)$ sectors of the theory. As we have seen above these
operators correspond to the special cohomologies of the $N=2$ string
in the $(-1,0)$ picture. Then we find $(n-1)$ physical states \phij\
corresponding to the deformations which respect the $N=4$ structure
(we show only left-moving index $m$ and omit $\bar{m}$)
\eqn\physsl{V_h=\CV_{n/2-h,-n/2+h}\CVp_{h,h}~,} where $h$ runs over $n-1$
half integers $h=1,3/2,....,n/2$.

One also finds $(n-1)$ antichiral states, which correspond to
$(0,-1)$ picture special cohomologies of the $N=2$ string
\eqn\physslsf{J^{--}_0V_h=\CV^{(s=-2)}_{n/2-h,-n/2+h}\CVp^{(s=-2)}_{h,h}=
\CV_{h-1,h-1}\CVp_{n/2-h+1,-n/2+h-1}~,} where to get the last line we
used \eqivsu\ and \eqivsl. We will see below that they are needed to
define the correlators of the $N=4$ TST. These chiral and
anti-chiral states correspond to the two types of the deformations
$t^{(i)}_1$ and $t^{(i)}_2$ in \defint, respectively.

Finally we would like to mention again that there are other physical
states \gp\ whose geometrical meaning is not clear. It would be an
interesting future problem to study them further. See \Ta\ for the
tree level scattering amplitudes for these states.

\newsec{Scattering Amplitudes in the $N=4$ Topological String on ALE}

\subsec{$N=4$ Topological String Amplitudes}

Let us consider $N\geq 4$ particle scattering amplitudes $A_N$ in $N=4$
topological string at tree level. Following Berkovits and Vafa \BV\
it is defined by 
\eqn\correlation{A_N=\left\la \left[\int d^2z_1 J_L^{--}J^{--}_R
V_1(z_1)\right]\ V_2(z_2) \ V_3(z_3)\ V_4(z_4) \prod_{a=5}^{N}\int
d^2 z_a G_L^{-}G_R^{-}V_a(z_a)\right\lb_{TST}.} The physical states
$V_i$ have the R-charge $R=1$ and the topological dimension $0$. The
total R-charge is $\sum_{i=1}^{N}q_i -2-(N-4)=2=\hat{c}$, which
shows the charge conservation violation (or ghost number anomaly)
familiar in topological string.

The integrated vertex operators $\int d^2 z_a
G_L^{-}G_R^{-}V_a(z_a)$ correspond to one of the four deformations
$t_{1L}t_{1R}$ in
\defint. The other three vertex operators can be found by replacing
either or both of $G^-_{L,R}$ with $\ti{G}^-_{L,R}$. 

To compute the correlation function \correlation\
we need to rewrite it in terms of untwisted correlators.
Then we can compute it using the results in conformal
field theory. This can be done by inserting the spectral flow
operator $J^{--}$. We insert it at the point $z=z_2$, though it can be inserted at any point
(refer to e.g. \Wa). Then the
\correlation\ can be written as
\eqn\correlationsp{\eqalign{A_N&=|z_2-z_3|^2|z_2-z_4|^2 \Bigl\la
\left[\int d^2 z_1 |z_1-z_2|^{-2}
J_{0L}^{--}J_{0R}^{--}V_1(z_1)\right]\ [J_{0L}^{--}J^{--}_{0R}
V_2(z_2)]\cr & \ \ \ \ \ \ \ \ \cdot V_3(z_3)\ V_4(z_4)
\prod_{a=5}^{N}\int d^2z_a G_{-1/2 L}^{-}G_{-1/2 R}^{-}V_a(z_a)
\Bigr\lb_{untwitsed}.} } 
This formula can be easily understood from
the point of view of the $N=2$ string. Indeed using \cc, \intgroo\
and \ingrfull\ we find 
\eqn\ntwocorr{ A_N=\langle
\phi_{1,int}^{(0,-1)}\phi_2^{(0,-1)}(z_2)
\phi_3^{(-1,0)}(z_3)\phi_4^{(-1,0)}(z_4)
\prod_{a=5}^N|PCO^-|^2\phi_{a,int}^{(-1,0)}\rangle~, } where
\eqn\phidefn{ \eqalign{ &\phi^{(-1,0)}=ce^{-\phi_+}V~,\cr
&\phi^{(0,-1)}=(S^+)^2\phi^{(-1,0)}~, }} and the operator
$\phi_{int}$ is defined by formulae \intgroo\ and \intgroopico. 
We see
that only the first term in \ingrfull\ contributes to the correlator
\ntwocorr\ due to anomalous conservation of $\p\phi_\pm$ currents.

{}From the physical string theory viewpoint, the amplitudes $A_N$
\correlationsp\ or \ntwocorr\ compute the coupling of  
4 RR-fields and $N-4$ NSNS-fields in the Little String
Theory\foot{In terms of the low-energy effective action, these
interactions schematically look like $\Tr[F^4 B^{N-4}]$, where
$F$ is the six dimensional $U(1)^n$ gauge field  and
$B$ is the Higgs boson corresponding to the transverse motion of
NS5-branes.}. In the case of the four point functions this was
discussed in \AFKS. 

We can identify the vertex $V_a$ and the spectral flowed one
$J^{--}_{0L}J^{--}_{0R}V_a$ in \correlationsp\ with the physical
states \physsl\ and \physslsf. In particular, we will find
convenient the following expressions which are equivalent to
\physsl\ and \physslsf\ via the identity \chiralsu (we suppress the
right-moving part) \eqn\revertr{\eqalign{ V_{h}&=
\CV^{(s=2)}_{h-1,h-1}\CVp^{(s=0)}_{h,h}~,\cr
J^{--}_{0L}J^{--}_{0R}V_{h}&=\CV^{(s=0)}_{h-1,h-1}\CVp^{(s=-2)}_{h,h}~.
}} We also find another equivalent representation using \eqivsl\
\eqn\anoth{ \eqalign{ V_{h}&=\CV^{(s=0)}_{n/2-h,-n/2+h}
\CVp^{(s=2)}_{\f{n+2}{2}-h,-\f{n+2}{2}+h}~, \cr
J^{--}_{0L}J^{--}_{0R}V_{h}&=\CV^{(s=-2)}_{n/2-h,-n/2+h}
\CVp^{(s=0)}_{\f{n+2}{2}-h,-\f{n+2}{2}+h}~. }}

The integrated vertex operators\foot{We determined the normalization
of these operators such that the cyclicity of amplitudes holds.}
take the form $G_{-1/2}^{-}V$. The $G^-$ action is divided
into two parts depending on whether it acts on the $SL(2,R)$ or
$SU(2)$ part. The former is given by
\eqn\gmact{G^{-(sl)}_{-1/2}V_{h}=\CV^{(s=2)}_{h-1,h-1}\CVp^{(s=-2)}_{h,h+1}~,}
while the latter is \eqn\gmactsl{
G^{-(su)}_{-1/2}V_{h}=\CV^{(s=-2)}_{\f n2-h ,h-\f n2+1}
\CVp^{(s=2)}_{\f{n+2}{2}-h,-\f{n+2}{2}+h}~.}

We also need the vertex operators in which 
$G^-$ is replaced by $\ti{G}^-$. 
Since $V_{h}$ is chiral separately in both
components it is very easy to find the action of $\ti G^-_{-1/2}$
\eqn\tiG{ \ti G^-_{-1/2}V_{h}=[J_0^{--},G^+_{-1/2}]
V_{\theta=0}=-G^+_{-1/2}J_0^{--}V_{j}~. } This is divided into two
parts: one obtained from the $SU(2)$ action of $\ti{G}$
\eqn\suact{\ti{G}^{-(su)}_{-1/2}V_{h}=-
\CV^{(s=2)}_{h-1,h-2}\CVp^{(s=-2)}_{h,h}~.} and the other one from
the $SL(2,R)$ action \eqn\slact{\ti{G}^{-(sl)}_{-1/2}V_{h}=-
\CV^{(s=-2)}_{\f{n}{2}-h,-\f{n}{2}+h}
\CVp^{(s=2)}_{\f{n+2}{2}-h,-\f{n+2}{2}-1+h}~.}

The correlators which involve $\ti G^-$ can be treated in the
$N=2$ string equally well. In general if one has an amplitude with
$L-4$ $G^-$ and $N-L$ $\ti G^-$, then
\eqn\ntwocorrNL{ A_{(N,L)}=\langle
\phi_{1,int}^{(0,-1)}\phi_2^{(0,-1)}(z_2)
\phi_3^{(-1,0)}(z_3)\phi_4^{(-1,0)}(z_4)
\prod_{a=5}^L|PCO^-|^2\phi_{a,int}^{(-1,0)}\prod_{a=L+1}^N
|PCO^+|^2\phi_{a,int}^{(0,-1)}\rangle~,}

\subsec{Relation between $SL(2,R)$ WZW Model and Bosonic Liouville
Theory}

To evaluate \correlationsp, we need the $N$-point correlation
functions in the supercoset SCFT \cosetdefg. They are essentially
reduced to the bosonic $SU(2)_{n-2}$ and $SL(2,R)_{n+2}$ WZW model
as can be seen from the formulae \pftosu\ and \pftosl. This is
because the fermions (i.e. super-partners)in the supercoset are
essentially free.

Even though there are no known general expressions for them except
for the three point functions \refs{\Tesch, \Hos}, recently a remarkable
relation has been uncovered \refs{\Sto, \RT, \Ri, \GN} between
correlation functions in the bosonic $SL(2,R)_{n+2}$ WZW
model\foot{We assume the usual analytical continuation of the
$H_3^+$ model to find results in the $SL(2,R)$ WZW model.}
 and those in
the bosonic Liouville theory. In this subsection we review this
relation for later convenience. In the recent paper \NN, the
computations of $N$-point functions in the $N=2$ topological string
on $SL(2,R)/U(1)$ WZW model have been done in order to find further
evidence for the equivalence between the twisted coset with the
level $n=1$ or $n>1$ and the $c=1$ string \MKV\ or the (non-minimal)
$c<1$ string \TMV (see also the recent discussion \AMT).

The bosonic Liouville field $\phi$ has background charge $Q=b+1/b$ and the
Liouville interaction is given by
\eqn\liouv{{\cal L}_{int}=\mu\int d^2 z e^{2b\phi}~.}
In the mentioned
relation, the bosonic $SL(2,R)_{n+2}$ model is mapped to the bosonic
Liouville theory with $b=\f{1}{\s{n}}$ and $\mu=\f{b^2}{\pi^2}$. This
model has the central charge $c=1+6(n+1)^2/n$. Notice that when $n$
is integer this Liouville theory appears in the $(1,n)$ minimal
bosonic string. The (spectral flowed) primary field
$\Phi^{w}_{h,m,\bar{m}}$ is mapped to the primary
$U_{\gamma}=e^{2\gamma\phi}$ in the Liouville CFT, where $\gamma$ is
defined by\foot{Notice that our definition of $h$ is related to the
ordinary spin $j$ in \refs{\Sto, \RT, \Ri, \GN} via $j=-h$.}
\eqn\alpj{\gamma=b(1-h)+\f{1}{2b}=\f{1}{\s{n}}(1-h)+\f{\s{n}}{2}~.}
Then the explicit map of $N$-point functions in both theories is
given by\foot{Our definition of the winding number $w$ is opposite
to the references mentioned. In terms of the primary in the $N=2$
coset, the quantum number $s$ is related to $w$ via \betsl \wnds.}
\refs{\Sto, \RT, \Ri, \GN} \eqn\srtmap{\eqalign{&\la
\prod_{i=1}^N\Phi^{w_i}_{h_i,m_i,\bar{m}_i}(z_i)\lb_{\sum_{i}w_i=r>0}
=\f{2\pi^{3-2N}b(c_{n+2})^r}{(N-r-2)!}\prod_{l=1}^N
N_{h_l,m_l,\bar{m}_l}\cdot
\delta^{(2)}\left(\sum_{l}m_l-\f{n+2}{2}~r\right)\cr & \ \ \ \
\times |\prod_{l<l'}z^{\beta_{ll'}}_{ll'}|^2\int
\prod_{a=1}^{N-2-r}dy^2_a \prod_{a<a'}|y_{a}-y_{a'}|^{n+2}\cdot
|\prod_{l,a}(z_l-y_a)^{-\f{n+2}{2}+m_l}|^2\cr& \ \ \ \ \times
\la\prod_{l=1}^N U_{\ap_l}(z_l)
\prod_{a=1}^{N-2-r}U_{-\f{1}{2b}}(y_a)\lb~,}} where we defined
$z_{ij}=z_i-z_j$ and \eqn\btnleg{\eqalign{\beta_{ll'}&=\f{n+2}{2}
-\f{n+2}{2}w_lw_{l'}+w_lm_{l'}+w_{l'}m_{l} -m_l-m_{l'}~,\cr
N_{h,m,\bar{m}}&=\f{\Gamma(h-m)}{\Gamma(1+\bar{m}-h)}~.}} Also
$c_{n+2}$ is a certain unknown constant which depends on $n$. A
similar formula for total negative winding number $\sum_{i}w_i=-r<0$
can be found by setting $m_i\to -m_i$.

Below we will be interested in the case $r=N-2$, where the maximal
winding number violation occurs. Only in this case (and also in the
minimally violated case), we do not have any insertions of the
vertex $U_{-\f{1}{2b}}(y_a)$ and get the following simple
formula\foot{This relation can be intuitively understood from the
equivalence between the $N=2$ twisted coset $SL(2,R)_{n+2}/U(1)$ and
the $c\leq 1$ string as noted in \refs{\TMV,\NN}. Consider the free
field representation (Wakimoto representation) of $SL(2,R)_{n+2}$
WZW model in terms of a bosonic scalar field with the background
charge $Q=\f{1}{\s{n}}$ as well as the bosonic ghosts
$(\beta,\gamma)$. After the topological twist, the background charge
of the scalar field becomes $Q=\s{n}+\f{1}{\s{n}}$, which is the
same as the one in the Liouville theory with $b=\f{1}{\s{n}}$. Then
we can indeed confirm the equivalence in the free field
representation as in \refs{\MKV, \TMV}.}
\eqn\srtmapmax{\eqalign{&\la
\prod_{i=1}^N\Phi^{w_i}_{h_i,m_i,\bar{m}_i}(z_i)\lb_{\sum_{i}w_i=N-2}
=2\pi^{-1}b\cdot(c_{n+2}/\pi^2)^{N-2}\cdot\prod_{l=1}^N
N_{h_l,m_l,\bar{m}_l}\cr &\ \ \cdot
\delta^{(2)}\Bigl(\sum_{l}m_l-\f{n+2}{2}(N-2)\Bigr)\cdot
|\prod_{l<l'}z^{\beta_{ll'}}_{ll'}|^2\cdot \la\prod_{l=1}^N
U_{\ap_l}(z_l)\lb~.}}

\subsec{Emergence of $(1,n)$ Minimal Bosonic String}

To analyze \correlationsp\ we also need to express the $N$-point
functions of the bosonic $SU(2)_{n-2}$ WZW in terms of correlators
in the minimal model. In order to establish this correspondence
recall that the $SU(2)$ algebra at the level $k=n-2$ is the same as
the  $SL(2,R)$ algebra at the negative level $k=-n-2$. Employing
this fact, we can regard the correlation functions of the $SU(2)$
model as those in the $SL(2,R)$ model.

This leads to the relation between the bosonic $SU(2)_{n-2}$ WZW
model and the bosonic Liouville theory with the imaginary parameter
$b=-\f{i}{\s{n}}$, which has the central charge $c=1-6(n-1)^2/n$.
The (spectral flowed) primary fields $\Psi^{w}_{j,m}$ 
are mapped to the $U_{\gamma'}=e^{2i\gamma'\vp}$ in the Liouville
theory, with  
\eqn\mapminma{\gamma'=-\f{1}{\s{n}}\left(j+1\right)+\f{\s{n}}{2}~.}
The map between correlation functions 
can be obtained from \srtmap, \srtmapmax\ by taking  $n\to -n$ and
$j\to -h$.

After the Wick rotation, this theory becomes the time-like Liouville
theory \refs{\ST, \SC}. It has the same central charge as the
minimal $(1,n)$ model. Thus we can expect that the correlation
functions in this Liouville theory and the minimal $(1,n)$ model are
the same. For $(p,q)\ \ (1<p<q)$ models, this equivalence was
checked in \Zamo\ by computing the three point functions. 

The minimal $(1,n)$ model can not be regarded as a minimal model in
a usual sense because its  Kac table is empty (there are only
$(p-1)(q-1)/2$ primaries in the Kac table of minimal $(p,q)$ model).
However, it is known that such a matter CFT coupled with
the Liouville theory is a well-defined string theory and it plays an
important role especially from the matrix model viewpoint (see e.g.
the review \TopR). 
In particular, one can deform the $(1,n)$ minimal string
into the $(p,n)$ minimal string as can be
understood from the integrable hierarchy arguments.

There are infinitely many physical states in the $(1,n)$ minimal
string. In the Coulomb gas representation they are given by
\eqn\primarymin{c T_{(r,s)}=c W_{(r,s)}\cdot
e^{\f{r+1-(s-1)n}{\s{n}}\phi}\ \ \ (r=1,2,...,n-1, \ \
s=1,2,3,....)~.}
The $r$-quantum number is truncated as usual by the screening charge, 
while $s$ is unrestricted.
(see \refs{\BNW, \GRK, \RW} for more
details). The operators $W_{(r,s)}$ are the primary fields 
in the $(1,n)$ model
\eqn\prmin{W_{(r,s)}=e^{2i\ap_{r,s}\vp},\ \ \
\ap_{r,s}=-\f{1-r}{2}\f{1}{\s{n}}+\f{1-s}{2}\s{n}~.}

The particular operators $T_{(r,1)}$ (i.e. $s=1$) will play an
important role in the later discussions. In the context of
topological gravity (see the appendix B), they correspond to the
matter chiral states while the others $s=2,3,\ddd$ are the
gravitational descendants. The lowest one $T_{(1,1)}$ is called the
puncture operator.

If we combine results of the previous and present subsections, we can
find an interesting connection between the supercoset \cosetdefg\
and the $(1,n)$ minimal string. Indeed we found that the $SL(2,R)$
part is mapped to the bosonic Liouville theory ($c_L=1+6(n+1)^2/n$),
while the $SU(2)$ part is mapped to the $(1,n)$ model ($c_m=1-6(n-1)^2/n$).
Together they give the matter part of the minimal $(1,n)$ model (note that
$c_m+c_L=26$). This motivates us to
investigate the correlation functions further in order to see if
those two theories are related.

\subsec{Four Point Functions}

Now we move back to the analysis of the scattering amplitudes
\correlation. Let us begin with the four point functions. By
applying the formula \srtmap, we can rewrite the four point function
for the vertices \revertr. Since the total winding number is
$\sum_{i=1}^4 w_i(=-2\sum_{i}s_i)=2$ in the $SL(2,R)$ sector, it is
maximally winding number violating amplitude. On the other hand, in
the $SU(2)$ sector, it corresponds to minimal winding number
violation. Then we can employ the simplified formula \srtmapmax\ and
its counterpart on the $SU(2)$ side. 

In addition to the four point function in the bosonic Liouville and
the $(1,n)$ model, we encounter some complicated factors of the form
$\prod_{l<l'}|z_{ll'}^{\beta_{ll'}}|^2$ when we evaluate
\correlationsp. First of all, such factors arise in the formula
\srtmap\ in both $SU(2)$ and $SL(2)$ sectors. Also we have to take
into account similar factors \corfreeone\ and \corfreetwo\ , which
arise when we relate the correlation functions of the $N=2$ cosets
to the original WZW models via \pftosu\ and \pftosl. Interestingly,
we find that all these factors are almost canceled with each other,
leaving us the simple factors $|z_1-z_2|^2 |z_3-z_4|^2$. Combined
with the similar factors in the original expression \correlationsp\
we obtain $|z_2-z_3|^2 |z_2-z_4|^2 |z_3-z_4|^2$ in the end. Notice
that this is the same as the familiar $c$-ghost correlation function
$\la c(z_2)\bar{c}(z_2)c(z_3)\bar{c}(z_3)c(z_4)\bar{c}(z_4)\lb$.

In this way we can relate the four point functions in $N=4$ TST to
the ones in $(1,n)$ minimal bosonic string (refer to \refs{\GLK,
\DK} for the correlation functions in minimal bosonic string) as
follows \eqn\corfour{\eqalign{A_4&=C\cdot
\delta^{(4)}(\sum_{i=4}^4h_i-2-n)\cdot \int d^2 z_1\la
c(z_2)\bar{c}(z_2)c(z_3)\bar{c}(z_3)c(z_4)\bar{c}(z_4)\lb \cr & \ \
\times  \left\la T_{(r_1,1)}(z_1,\bar{z}_1)~
T_{(r_2,1)}(z_2,\bar{z}_2)~T_{(r_3,1)}(z_3,\bar{z}_3)
~T_{(r_4,1)}(z_4,\bar{z}_4)\right\lb~,}} where the fields $T_{(r,1)}$
are the physical vertex operators in $(1,n)$ bosonic string, defined
in \primarymin, and they correspond to the operators
$V_{h=1,3/2,\ddd,n/2}$ in \revertr\ via the relation
\eqn\relationrr{ r=n+1-2h\ \ \ (=1,2,...,n-1)~.} This relation
between $h$ and $r$ comes from the maps \alpj\ and \mapminma. The
factor $C$ is an overall constant $C=\f{-2i}{\pi^2 n}\cdot
(\f{c_{n+2}c_{-n+2}}{\pi^{2}})^2\left(\f{\Gamma(0)}{\Gamma(1)}\right)^8$;
we can absorb the divergent piece\foot{We can also absorb the factor
$c_{n+2}c_{-n+2}\pi^{-2}$ in the normalization. This is because
 the $N$-point functions
always include the factor $(c_{n+2}c_{-n+2}\pi^{-2})^{N-2}$. }
 $(\f{\Gamma(0)^2}{\Gamma(1)^2})^N$ in
the normalization of each vertex $V_{h_i}$ in the $N=4$ TST.

We have found that the physical states $V_{h=1,3/2,...,n/2}$
\physsl\ in the $N=4$ topological string on ALE spaces are in one to
one correspondence with the states $T_{(r=1,2,..,n-1, \ s=1)}$ in
the minimal $(1,n)$ string. It is also natural to expect that the
other states $T_{(r,s>1)}$ may correspond to some other physical
states in the $N=4$ topological string in a way similar to the
gravitational descendants, though we will not discuss this issue in
this paper.

Then the four point functions can be written simply as follows (up
to an unimportant factor and the delta function in the \relationrr)
\eqn\fourpps{A_4=\la T_{(r_1,1)}T_{(r_2,1)}T_{(r_3,1)}T_{(r_4,1)}\lb
_{(1,n)\rm {string}}~.} In addition we have the following
constraint from the R-charge conservation (i.e. the
$\delta$-function in \corfour) \eqn\rchargec{\sum_{i=1}^4 h_i=2+n,\
\ \ ({\rm{or}}\ \ {\rm{equivalently}} \ \sum_{i=1}^4r_i=2n)~. } This
constraint is not clear from the viewpoint of the $(1,n)$ minimal
string.

  On the other hand, if we apply the other equivalent representation
\anoth, then we find another expression after a similar analysis
\eqn\fourppss{A_4=\la T_{(\ti{r}_1,1)}\ T_{(\ti{r}_2,1)}\
T_{(\ti{r}_3,1)}\ T_{(\ti{r}_4,1)}\lb _{(1,n) string}~,} with the
same constraint \rchargec. Here we defined the integer $\ti{r}$ by
\eqn\deptrr{\ti{r}\equiv n-r=2h-1\ \ \ \ (=1,2,...,n-1)~.}

These two expressions \fourpps\ and \fourppss\ should be identical.
This suggests that the theory has the $\IZ_2$ global symmetry which
replaces all of the operators $T_{(r,1)}$ with the 
$T_{(\ti{r},1)}$. This symmetry is consistent with the four point
function expression\foot{It is intriguing to note that
 this expression coincides with the four point
function \WZW\ in the topological gravity with $(n-1)$-th minimal
matter. The latter theory is usually associated with $(1,n+1)$
bosonic string without Liouville potential as reviewed in
appendix B, instead of $(1,n)$
minimal bosonic string with Liouville wall. The
possibility of the connection between the $N=4$ TST and the
$(1,n+1)$ string has already been implied in \OV\ from the analysis
of R-charge conservation. For more details see the appendix B.}
conjectured in \AFKS\ from the duality between Heterotic string on
$T^4$ and Type II string on K3 \eqn\fourtpg{A_{4}=\min
\{r_i,n-r_i\}~.}

\subsec{Five Point Functions}

Now let us proceed to five point functions. In this case we can
still rewrite the correlation functions in terms of those in the
$(1,n)$ minimal string. We have two choices for the integrated
operator corresponding to the action of $G^-$ and $\ti{G}^{-}$. We
can again apply the formula \srtmapmax\ and simplify the total
expression in the same way as in the four point function case. In
the end, we obtain the following results classified into four cases
\eqn\fivepps{\eqalign{A^{(1)}_5&= \left\la \left(\int\
J^{--}_{0R}J^{--}_{0L}V_{h_1}\right)(J^{--}_{0R}J^{--}_{0L}V_{h_2})\cdot
V_{h_3}\cdot V_{h_4}\cdot \left(\int
G^-_{-1/2R}G^-_{-1/2L}V_{h_5}\right)\right\lb \cr &= \la
T_{(r_1,1)}T_{(r_2,1)}T_{(r_3,1)}T_{(r_4,1)}T_{(r_5,1)}\lb
_{\rm(1,n) string}\ \ \ \ {\rm when}\ \  \sum_i h_i=2+3n/2~, \cr
A_5^{(2)}&=\left\la \left(\int\
J^{--}_{0R}J^{--}_{0L}V_{h_1}\right)(J^{--}_{0R}J^{--}_{0L}V_{h_2})\cdot
V_{h_3}\cdot V_{h_4}\cdot \left(\int
G^-_{-1/2R}G^-_{-1/2L}V_{h_5}\right)\right\lb \cr &= \la
T_{(\ti{r}_1,1)}T_{(\ti{r}_2,1)}
T_{(\ti{r}_3,1)}T_{(\ti{r}_4,1)}T_{(\ti{r}_5,1)}\lb _{\rm(1,n)
string}\ \ \ \ {\rm when}\ \  \sum_i h_i=2+n~, \cr
 A_5^{(3)}&=\left\la \left(\int\
J^{--}_{0R}J^{--}_{0L}V_{h_1}\right)(J^{--}_{0R}J^{--}_{0L}V_{h_2})\cdot
V_{h_3}\cdot V_{h_4}\cdot \left(\int
\ti{G}^-_{-1/2R}\ti{G}^-_{-1/2L}V_{h_5}\right)\right\lb \cr &= \la
T_{(r_1,1)}T_{(r_2,1)} T_{(r_3,1)}T_{(r_4,1)}T_{(r_5,1)}\lb
_{\rm(1,n) string}\ \ \ \ {\rm when}\ \  \sum_i h_i=3+3n/2~, \cr
A_5^{(4)}&= \left\la \left(\int\
J^{--}_{0R}J^{--}_{0L}V_{h_1}\right)(J^{--}_{0R}J^{--}_{0L}V_{h_2})\cdot
V_{h_3}\cdot V_{h_4}\cdot \left(\int
\ti{G}^-_{-1/2R}\ti{G}^-_{-1/2L}V_{h_5}\right)\right\lb \cr &= \la
T_{(\ti{r}_1,1)}T_{(\ti{r}_2,1)}
T_{(\ti{r}_3,1)}T_{(\ti{r}_4,1)}T_{(\ti{r}_5,1)}\lb _{\rm(1,n)
string}\ \ \ \ {\rm when}\ \  \sum_i h_i=3+n~,}} where again we defined
$\ti{r}=n-r$. The above results for $A^{(1)}$ and $A^{(3)}$ are
found by using the expressions \revertr \gmact \suact, while those
for $A^{(2)}$ and $A^{(4)}$ are done by applying \anoth \gmactsl
\slact.

\subsec{$N(\geq 6)$-point Functions}

In the case of $N(\geq 6)$-point functions, the amplitudes
\correlationsp\  vanish unless the following R-charge
conservation is satisfied \eqn\rcgarff{\sum_{i=1}^N
h_i=n+2+l_1+\f{n}{2}\ l_2~,} where the integers $l_1$ and $l_2$ takes
the values $l_{1,2}=0,1,2,...,(N-4)$. If the numbers of insertions
of $G^{-(sl)}$, $G^{-(su)}$, $\ti{G}^{-(su)}$ and $\ti{G}^{-(sl)}$
are denoted by $a,b,c$ and $d$, then the integers $l_1$ and $l_2$
are written as $l_1=N-4-a-b$ and $l_2=a+c$.

We can again rewrite the $N$-point functions in terms of the
correlation functions in the bosonic Liouville theory using \srtmap.
Unlike the four and five-point functions, it is not obvious how to
reduce the general $N$-point functions to those in the $(1,n)$
string as long as we proceed just as before. This is because we do
not have the maximally or minimally winding violation $\sum_i
w_i=\pm (N-2)$ in general and we cannot eliminate the insertions of
$V_{-\f{1}{2b}}(y_a)$ in \srtmap.

However, we can find that some of the amplitudes can be rewritten in
terms of the $(1,n)$ string by applying \srtmapmax. Consider the
$N-$point function $A^{(1)}_N$ which includes $(N-4-l_1)$ $\
G^{-(sl)}V_h$ operators and $l_1$ $\  \ti{G}^{-(su)}V_h$ ones, and
also its dual amplitude $A^{(2)}_N$ obtained via $r \to \ti{r}$. We
can show\foot{This consideration also determines the normalization
of each vertex $V_{a}$ in \correlationsp. We rescale the operators
$V_{1,2,3,4}$ by multiplying the factor
$\pi^2(c_{n+2}c_{-n+2})^{-1}\Gamma(0)^{-2}$. For the other
integrated vertex operators, in addition to this factor, we also
need to multiply $i$ for each $G^{-}$ action and $-i$ for
$\ti{G}^{-}$ action.} that they
are the same as the $(1,n)$ minimal string amplitudes
\eqn\minampi{\eqalign{A^{(1)}_{N}&=\la
T_{(r_1,1)}T_{(r_2,1)}...T_{(r_N,1)}\lb_{(1,n){\rm  string}}\ \ \ \
{\rm when} \ \ \ \sum_{i}h_i=(N-2)\f{n}{2}+2+l_1~,\cr
A^{(2)}_{N}&=\la
T_{(\ti{r}_1,1)}T_{(\ti{r}_2,1)}...T_{(\ti{r}_N,1)}\lb_{(1,n){\rm
string}}\ \ \ \ {\rm when}\ \ \sum_{i}h_i=n+2+l_1~,}} where
$l_1=0,1,...,N-4$. These correspond to $l_2=N-4$ and $l_2=0$,
respectively in \rcgarff.

Naively, amplitudes other than \minampi\  do not seem to be
reduced to the ones in the $(1,n)$ string. 
If we apply \srtmap\ and rewrite
them, then they will include the extra insertions of integrated operators
of the form $\int dy V_{-1/2b}(y)$. 
However, we cannot deny the possibility that one can still relate the generic $N\geq 6$
amplitudes to the $(1,n)$ string amplitudes by performing the integral explicitly.
This point needs future investigations and we will not pursue it
here.

\subsec{$N=4$ Topological String on $A_1$ Space and Matrix Model for
$(1,2)$ String}

It is well-known that the $(1,n)$ minimal string is equivalent to
the double scaled multi-matrix model. In the simplest case of the
$(1,2)$ string ($c=-2$ string), it can also be thought of as the
$k=1$ case of the $(2,2k-1)$ series. Then its matrix model dual is
simply given by the gaussian one matrix model \bosmin. Since this matrix model
can be solved exactly, we would like to use it to obtain the
scattering amplitudes in $N=4$ topological string on the $A_1$ type
ALE space (Eguchi-Hanson space).

In this case the tree level correlation functions look
like\foot{Only in this subsection we recover the dependence on the
cosmological constant to make things clear.} \refs{\bosmin, \KW}
\eqn\corctwo{\eqalign{\la \prod_{i=1}^{N}T_{(1,s_i)}\lb &= \mu^{\sum
s_i-2N+3}\f{\Gamma(\sum_i s_i-N+1)}{\Gamma(\sum_i s_i-2N+4)}\cdot
\log\mu,\ \ \ \ \ (\sum_{i}s_i-2N+3\geq 0)\cr &=\mu^{\sum
s_i-2N+3}\Gamma(\sum_i s_i-N+1)\Gamma(2N-3-\sum_i s_i),\ \ \ \ \
(\sum_{i}s_i-2N+3<0)~,}} where we assumed $\sum_i s_i-N+1\geq 1$.
 Note that the terms with the positive powers of $\mu$ are
 accompanied with $\log\mu$, which is explained by the
infinite volume in the Liouville direction. 
Thus they survive the double scaling limit 
along with the terms with negative powers of $\mu$.


When we keep only the terms with $\log\mu$ singularity, the free
energy $F/\log\mu=t^3/6-1/12$ and the scattering amplitudes
\corctwo\ only include non-negative integer powers of the
cosmological constant $\mu$. In this sense the theory becomes
topological and it is known to be equivalent to the pure topological
gravity \WiT. The similar model in the $(1,n)$ case is also known to
be the same as the topological gravity coupled to the  $(n-2)$-nd
minimal topological matter \refs{\EY, \KL, \WZW, \DW, \TopR}. Its
matrix model dual is given by the generalized Kontsevich model. In
the relation to the $N=4$ topological string we also need to keep
the terms with negative powers of $\mu$ so that the R-charge
conservation \rchargec\ or \rcgarff\ is satisfied.

Now let us consider the $N$-point functions of the $N=4$ topological string, 
which only  involve the integrated operators of the form 
 $\int G^{-}_{L}G^-_{R}V_h$. Since in the $n=2$ case, the $SU(2)$ sector
becomes trivial, we can only insert $G^{-(sl)}$ to obtain
non-zero result. Hence 
 all such
tree level correlation functions can be rewritten in terms of the amplitudes of the
$(1,2)$ minimal string just as we did for  $A^{(1)}_N$ in \minampi. Now we 
can use the matrix model result  \corctwo\ to compute these
$N-$point correlation functions 
 \eqn\corenfhet{A_N=\mu^{3-N}\cdot (N-4)!~.}

\newsec{Discussion: Toward Matrix Model for Minimal $N=2$ String}

In this paper we studied the $N=2$ minimal string and
its connection to the $(1,n)$ minimal bosonic
string. The $N=2$ minimal string is equivalent to the 
$N=4$ topological string near the ALE singularity.
We concentrated on a particular set of physical states 
which correspond to the K\"ahler and complex structure deformations of the
underlying geometry.
Then we found one to one
correspondence between them and physical states in the $(1,n)$
minimal string. It would be interesting to perform an exhaustive
cohomology analysis and see if they are indeed equivalent at the
free theory level.

To investigate this relation at the interaction level, we computed
the closed string scattering amplitudes in the $N=2$ minimal string
or equivalently in the $N=4$ topological string on ALE spaces. Indeed
we found an intriguing connection to the $(1,n)$ minimal string. In
particular, we showed that all four and five-point functions can be
rewritten in terms of those of the $(1,n)$ minimal bosonic string.
We were not able to match generic $N(\geq 6)$-point functions in the
$N=2$ string to that of the $(1,n)$ bosonic string although some
classes of the higher point amplitudes do match.

These results suggest that these two string theories are closely
related. 
On the other hand there also seem to be
important distinctions. In particular we encountered the R-charge
conservation like \rchargec, which is not easy to understand from
the viewpoint of $(1,n)$ string. Also methods used in this paper do
not allow to match {\it generic} six and higher point functions. It
would be interesting to understand whether these problems are
artefact of the method or they indicate the non-equivalence of the
two theories. Therefore it is very important to have future progress
on both sides. Below we would like to discuss a possible matrix
model dual for the $N=2$ minimal string inspired by this connection.

\subsec{ADE Matrix Model} It is very natural to expect that 
there exists a matrix model dual for the minimal $N=2$ string 
or equivalently for the $N=4$
topological string on ALE spaces, as is true both in the bosonic and
type 0 minimal string. We expect that the dual matrix model is
equivalent to an open string theory of infinitely many D0-branes in
that string theory. It is dual to the closed string theory via the
holography as was so in the two dimensional string theory \MV. Refer
to \refs{\GRT, \minimal} for recent discussions of open-closed duality for
$(1,n)$ minimal string in different contexts.

The relevant D-branes in our model should be the 
D2-branes wrapped on $n-1$  2-cycles in the $A_{n-1}$ ALE
space. Indeed one can construct corresponding $N=2$ supersymmetric
(B-type) boundary states\foot{Furthermore, one can show that the B or
A-type boundary states preserve $N=4$ boundary conformal symmetry in
$\hat{c}=2$ theory as is natural from the viewpoint of $N=4$
topological string.} \ESD. The open string theory is expected to be
described by a quiver-like theory with $n-1$ nodes and $n-2$ arrows.

We would like to point out that a possible matrix model dual of the
$N=2$ string on $ADE$ ALE spaces may be given by the Kostov's $ADE$
matrix model \ADE. It is defined by the following matrix model
action in the $A_{n-1}$ case \eqn\actimat{
S=\sum_{a=1}^{n-1}\Tr[U(\Phi^{(a)})+M^{(a)}\bar{M}^{(a)}]
+\sum_{a=1}^{n-2}[\bar{M}^{(a)}\Phi^{(a)}M^{(a)}+
\bar{M}^{(a)}M^{(a)}\Phi^{(a+1)}]~.} 
Generalization to $D_n$ or $E_n$ case is straightforward.
In the $A_{n-1}$ quiver, the
eigenvalues of the adjacency matrix $C_{a,b}$ take the values
$\beta_{(p)}=2\cos(\pi p)$, where $p$ runs the values
$p=\f{1}{n},\f{2}{n},\ddd,\f{n-1}{n}$. Correspondingly, the matrix
model is known to describe the non-critical bosonic string with the
matter central charge $c=1-\f{6p^2}{1-p}$ after a suitable double
scaling limit.
 The choice of $p$
corresponds to the choice of the background charge on the
world-sheet. This fact becomes clearer in the equivalent RSOS model
description \KP. When we choose the maximal value $p=\f{n-1}{n}$ we
have $c=1-6\f{(n-1)^2}{n}$, which is the same as the $(1,n)$ minimal
bosonic string\foot{We can get a similar range of matter central
charge for affine $\hat{A}_{2n-1}$ quiver matrix model. We believe
this corresponds to the non-minimal $(1,n)$ string which is
equivalent to the $N=2$ topological string on $SL(2,R)/U(1)$ \TMV.}.
Notice that in the simplest case of $n=2$, the matrix model
\actimat\ is reduced to the gaussian one matrix model, which is
known to be equivalent to the $(1,2)$ string.

The eigenvectors of $C_{a,b}$ are given by
$v^{(a)}_{p}=\s{\f{2}{n}}\sin(\pi pa)$. For each $p=\f{h}{n}\
(h=1,2,...,n-1)$, we have a corresponding operator which may be
identified with the closed string primary $T_{(r=h,1)}$ \primarymin.
The integer $h$ is the discrete Fourier transformation with respect
to the `position' $a=1,2,..,n-1$. Notice also that the $\IZ_2$
symmetry $T_{(r,1)}\lr T_{(\ti{r}=n-r,1)}$ mentioned in section 4.4
becomes obvious in this description, being identified with the
$\IZ_2$ reflection symmetry of $A_{n-1}$ Dynkin diagram.

\subsec{World-Sheet Discretization}

Another way to find a matrix model dual to closed string is 
via worldsheet discretization.
How should the candidate matrix model
discretize the worldsheet of the $N=4$ topological string 
embedded into the ALE space? Since the
string theory is topological, we expect that the worldsheet is
localized on the non-trivial two-cycles of the target space. 
Then the sigma model map from a Riemann
surface $\Sigma$ to the $A_{n-1}$ ALE space {\it is now reduced to
a map from $\Sigma$ to the $n-1$ points}. These $n-1$ points specify
which 2-cycle a point on the Riemann surface is situated at.
Hence the worldsheet is divided into regions, and 
the adjacent regions are mapped into the adjacent two cycles.

The ADE matrix model \actimat\ does precisely this as we explain below.
When we pick a Feynman diagram for \actimat, we find that the
chains of propagators of $M^{(a)}$ field behave as non-intersecting
loops. They divide the whole net of the Feynman diagram into
regions bounded by these loops (see the left figure in Fig. 1). Such a
model is known as the loop gas model (or $O(n)$ model). By taking
its dual lattice, it is also equivalent to the model called RSOS
model \KP.

{}From this viewpoint, we have infinitely many domains surrounded by
the loops on random Riemann surfaces. We can assign one number $a$
(called height variable) out of the $n-1$ integers $a=1,2,3,...,n-1$
to each domain. In the matrix model \actimat\ language, this is the
index $a$ of the matrices $\Phi^{(a)}$. Then we have the requirement
that the two domains which are adjacent should take the values $a$
which are different from each other by $\pm 1$ as is clear from the
matrix model. Finally we assign a specific weight for a loop and sum
over all such configurations. In this way we find that this model
describes the map from a random surface to $n-1$ points (specified
by the integer $a$) as we expected for the topological string on ALE
spaces (see the right figure in Fig.1). Notice that this integer $a$
describes the types of 2-cycles which D2-branes are wrapped on and
thus this matches the above observation. The origin of $N=2$
world-sheet supersymmetry is not clear from this observation,
unfortunately. This issue has not completely been understood even in
the type 0 string theory, though there have been some progress
\GRT. We leave further analysis for future publications.

\fig{Discretization of world-sheet and the sigma-model map into
$A_{7}$ ALE space.}{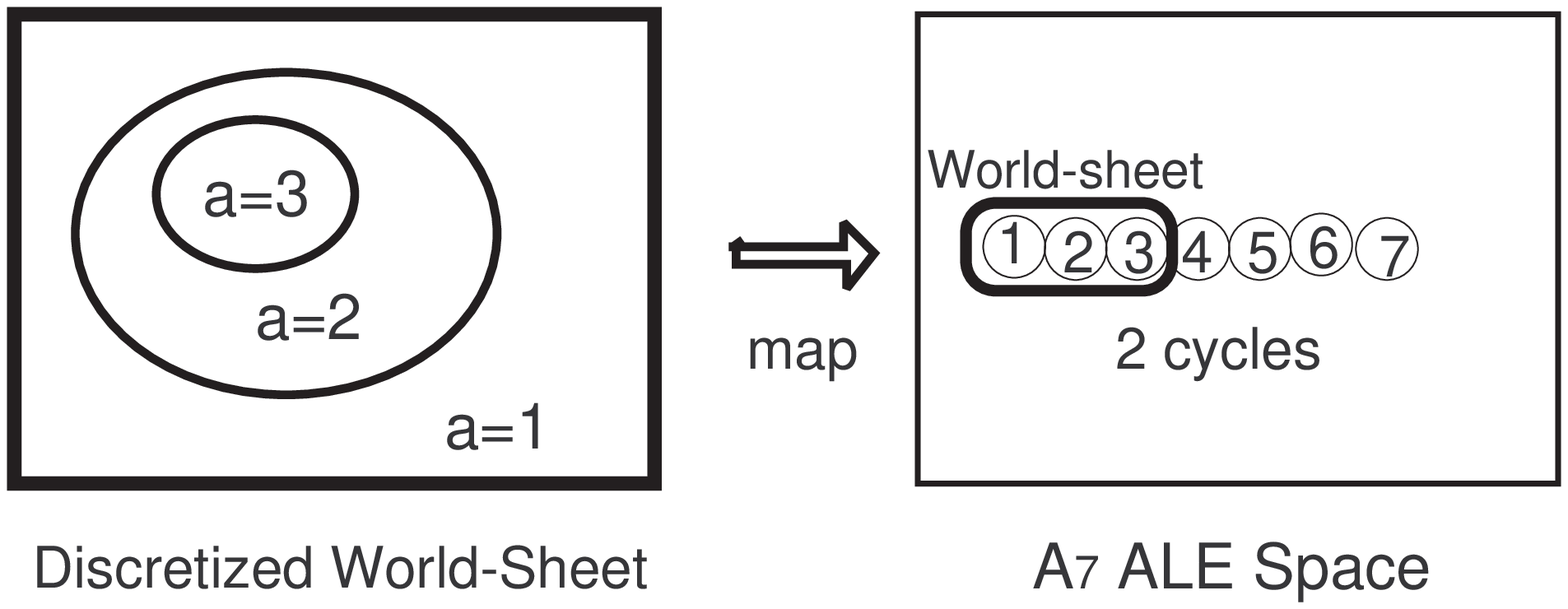}{4truein}

\vskip 0.2in

\centerline{\bf Acknowledgments}

We would like to thank S. Ashok, D. Gaiotto, S. Gukov,
K. Hosomichi,  A. Konechny, D. Kutasov, A. Parnachev, S. Ribault, S. Ryu, S.
Terashima, J. Teschner and C. Vafa for useful discussions and
correspondences. We are grateful to the KITP program ``Mathematical
Structures in String Theory" during which this work was initiated
and completed. This work was supported in part by the National
Science Foundation under Grant No.~PHY99-07949 and by the Department
of Energy under Grant DE-FG02-96ER40949.

\vskip 0.2in

\appendix{A}{$N=4$ Topological String on $\IC^2/\IZ_n$}

Here we study physical states of the $N=4$ topological string on
$\IC^2/\IZ_n$ (see also \GOSZ\ for $N=2$ string on $\IC^2/\IZ_2$).
The world-sheet theory is described by the scalar fields
$(X^1,X^2,\bar{X}^1,\bar{X}^2)$ and their superpartners
$(\psi^1,\psi^2,\bar{\psi}^1,\bar{\psi}^2)$. The $\IZ_n$ orbifold
action is defined by $(X^1,X^2)\to (e^{2\pi i/n}X^1, e^{-2\pi
i/n}X^2)$. The (left-moving) superconformal generators are
\eqn\scg{G^+=\de \bar{X}^1\psi^1+\de\bar{X}^2\psi^2,\ \ \ \
\ti{G}^+=\de X^1\psi^2-\de X^2\psi^1~.} In the $k-$th and $(n-k)$-th
twisted sector of the bosonic part of $\IZ_n$ orbifold, the ground
states are represented by the twist operators
$\sigma_{\pm}(z,\bar{z})$, which are defined by the OPEs $(i=1,2)$
\eqn\opeorbf{\eqalign{ &\de X^1(z)\sigma_+^1(0) \sim
z^{-1+k/n}\tau^1_+, \ \ \ \de \bar{X}^1(z)\sigma^1_+(0)\sim
z^{-k/n}\tau^1_{+'}~,\cr & \db X^1(\bar{z})\sigma_+^1(0) \sim
\bar{z}^{-k/n}\ti{\tau}^1_+, \ \ \ \db
\bar{X}^1(\bar{z})\sigma^1_+(0)\sim
\bar{z}^{-1+k/n}\ti{\tau}^1_{+'}~,\cr & \de X^1(z)\sigma^1_-(0) \sim
z^{-k/n}\tau^1_-, \ \ \ \de \bar{X}^1(z)\sigma_-^1(0)\sim
z^{-1+k/n}\tau^1_{-'}~, \cr & \db X^1(\bar{z})\sigma^1_-(0) \sim
\bar{z}^{-1+k/n}\ti{\tau}^1_-, \ \ \ \db
\bar{X}^1(\bar{z})\sigma_-^1(0)\sim
\bar{z}^{-k/n}\ti{\tau}^1_{-'}~.}} The similar OPE between $X^2$ and
$\sigma^2_{\pm}$ can be obtained by replacing $k/n$ with $1-k/n$ in
\opeorbf. Notice also that  if we go around $\sigma^1_{\pm}$ once,
we will get the twist by $e^{\pm 2\pi ik/n}$. Therefore, we are
indeed considering the $k-$th and $(n-k)-$th twisted sector at the
same time.

The condition \physco\ requires that the OPEs $G^+(z)V_{phys}(0)$,
$\ti{G}^+(z)V_{phys}(0)$,  $G^+(\bar{z})V_{phys}(0)$ and
$\ti{G}^+(\bar{z})V_{phys}(0)$ do not include the singularities
$z^{-n}$ or $\bar{z}^{-n}\ \ (n\geq 1)$. Then we can find only one
physical state in each $k-$th twisted sector \eqn\physnf{V^{N=4
TST(k)}_{phys}=\sigma^1_+\sigma^2_{+}e^{i\f{k}{n}H_1}e^{i(1-\f{k}{n})H_2}
e^{i(1-\f{k}{n})\ti{H}_1}e^{i\f{k}{n}\ti{H}_2}~,} where $H_{1,2}$ are
the bosonizations of the left-moving fermions
$\psi^{1,2}(z)=e^{iH_{1,2}}$ (OPEs are $H_i(z)H_j(0)\sim
-\delta_{ij}\log z$).  $\ti{H}_{1,2}$ are the ones for the
right-moving part.

Indeed it has the R-charges $J_3=\ti{J}_3=1$ and the conformal
dimension $\Delta=1/2$ before twisting (notice that
$\Delta(\sigma_{\pm})=-\f{1}{2}(\f{k}{n})^2+\f{1}{2}(\f{k}{n})$).
This operator \physnf\  is $(chiral, chiral)$ primary in the $N=2$
SCFT and also belongs to the $k-$th twisted sector. By replacing
$\sigma^{1,2}_+$ with $\sigma^{1,2}_{-}$, we find the physical state
for $(n-k)-$th twisted sector. In this way we found $n-1$ twisted
sector physical states. Geometrically they nicely correspond to the
$n-1$ blowing up modes in the $A_{n-1}$ ALE space. This number of
physical states agrees with the singular ALE model studied in
section 2, as we expected because these two different backgrounds
should correspond to two different points in the moduli space of
$A_{n-1}$ ALE space. Note also that in this model we do not have the
states written as $G^{+}_{-1/2}V$ opposite to the singular ALE case.
Equivalently there are no $(-1,-1)$ picture physical states in this
$N=2$ string except the trivial one.

\appendix{B}{Topological Gravity}

 Consider the twisted $p-$th $N=2$ minimal model
($c=\f{3p}{p+2}$) corresponding to the LG potential $W=X^{p+2}$ \EY.
We can use the twisted $N=2$ minimal model as a matter theory to
define the topological gravity \refs{\WZW, \DVVT, \DW} (see also the
excellent review \TopR\ on these matters). The most important
physical states are chiral primaries and can be expressed as
$\phi^{(m)}\sim X^m\ \ \ (m=0,1,...p)$. The lowest operator
$\phi^{(0)}$ is called the puncture operator. There are other types
of physical states called gravitational descendants, which we will
not consider here.

The state $\phi^{(m)}$ has the R-charge $q=\f{m}{p+2}$. The
$N-$point amplitudes are defined by \eqn\amptm{A_{N}=\la
\phi^{(i_1)}\phi^{(i_2)}\phi^{(i_3)}\prod_{a=4}^N\int
G_L^{-}G_R^{-}\phi^{(i_a)}\lb~.} The R-charge (or ghost charge)
conservation leads to
\eqn\ghostconr{\sum_{i=1}^N\f{m_i}{p+2}-(N-3)=\hat{c}=1-\f{2}{p+2}~.}

As we have mentioned in section 4, this theory of topological
gravity is known to be equivalent to the $(1,n)$ minimal bosonic
string \refs{\KL, \BNW}. In this context, the condition \amptm\ can be
understood by considering a minimal $(1,n)$ bosonic string {\it
without Liouville potential} (but with the screening charge in the
minimal matter sector). We identify $\phi^{(m)}$ with the tachyon
operator \eqn\identitt{T_{(r=n-1-m,s=1)}=W_{(r,1)}\cdot
e^{\f{(n-m)}{\s{n}}\phi}~,} setting $n=p+2$. Indeed the momentum
conservation in the Liouville $\phi$ direction coincides with
\ghostconr. Furthermore, we can even prove that all tree level
$N$-point functions of the $N=2$ twisted minimal model are the same
as those in the $(1,n)$ minimal string using the relation \srtmap.
Since this computation is analogous to what we did in section 4 for
the $SU(2)/U(1)$ sector, we omit the details.

In topological gravity, we usually talk about correlation functions
expanding around the point where the cosmological constant is
vanishing. Three and four point function \refs{\DVVT, \WZW} are given by
\eqn\corfff{\eqalign{ \la
\phi^{m_1}\phi^{m_2}\phi^{m_3}\lb&=\delta_{m_1+m_2+m_3,p}~, \cr \la
\phi^{m_1}\phi^{m_2}\phi^{m_3}\phi^{m_4}\lb&
=\f{1}{p+2}\min(m_i,p+1-m_i)\cdot
\delta_{m_1+m_2+m_3+m_4,2p+2}~.}} The delta-function in three point
function comes from the R-charge conservation as already mentioned.

Finally, let us compare this with $N=4$ TST. Looking at the $SU(2)$
part in our case, let us identify\foot{Notice that this is different
from more standard one \identitt.} $r=n-m$ assuming $p=n-1$. The
state $m=0$ is almost trivial as it vanishes when acted upon by
$G^-$ and thus it is plausible that it is absent in $N=4$ TST. Then
\ghostconr\ can be rewritten as $\sum_{i=1}^N r_i=2n+4-N$. This
agrees with the R-charge constraint of the four point function
$A^{(4)}$ and the special amplitudes $A^{(1)}_N$ at $l_1=N-4$ in
\rchargec\ \minampi. Also the four point function itself agrees (up
to a constant) with the four point function \corfff\ in $N=4$ TST
obtained from the Het/TypeII duality setting $n=p+1$ again. Even
though it is possible that this is just a coincidence, it would be
interesting to see if this is indeed true in general amplitudes.

\appendix{C}{Closed String Partition Functions}

Here we compute the partition function for the $N=2$ minimal string.
Define the NS-sector N=2 character for the $\hat{c}=2$ matter N=2
SCFT as usual
\eqn\defpart{Z_{\hat{c}=2}(\tau,z)=\Tr[q^{L_0-c/24}\bar{q}^{\bar{L}_0-c/24}
y^{J_0}\bar{y}^{\bar{J}_0}]~,} where $q=e^{2\pi \tau}$ and $y=e^{2\pi
iz}$. The expression of the $N=2$ string partition function is given
by  \refs{\ABK, \OV}  \eqn\defntpar{Z_{N=2}=\int \f{d\tau
d\bar{\tau}}{\tau_2}\int dz
d\bar{z}|Z_{\hat{c}=2}(\tau,z)|^2|Z_{gh}(\tau,z)|^2~,} where
$(\tau,\bar{\tau})$ denotes the ordinary torus moduli and
$(z,\bar{z})$ denote the $U(1)$ gauge field moduli peculiar to $N=2$
string; we parameterized $z=\theta_1+\tau\theta_2$ assuming
$0\leq\theta_{1,2}<1$.

The part $Z_{gh}(\tau,z)$ represents the ghost partition function
and it is explicitly expressed as
\eqn\ghostp{Z_{gh}(z|\tau)=\f{\eta(\tau)^6}{\theta_3(z|\tau)^2}~.}

Let us proceed in our specific example \cosetdefg, taking into
account the continuous modes and not the discrete states for
simplicity. We expect this is enough if we are interested in the
bulk terms which behave like $\sim \log\mu_{sl}$, where $\mu_{sl}$
is the $N=2$ cosmological constant as we usually do in the bosonic
or type 0 non-critical string to compare matrix models.

The final result reads \eqn\totalpp{\eqalign{Z_{N=2}= c\cdot
\log\mu_{sl} \int \f{d\tau d\bar{\tau}}{(\tau_2)^2}\int^{1}_0
d\theta_1d\theta_2\ (\tau_2)^{3/2}|\eta(\tau)|^6
\sum_{l,l'}N_{l,l'}\chi^{(l)}(\tau,0)\chi^{(l')}(\bar{\tau},0)~.}}
$\chi^{(l)}$ is the spin $j=l/2$ character of $SU(2)_{n-2}$ WZW
model. The $N_{l,l'}$ represents the multiplicity of primaries and
it is well-know that it has the ADE classification corresponding to
the ADE classification of ALE spaces \refs{\OV, \ABKS, \ESP}. $c$ is
a computable constant if we fix normalization. This expression
\totalpp\ is manifestly $N=2$ modular invariant. Notice that the
partition function in the end does not depend on $z$ as was also
true in the $N=2$ string on $\IR^{2,2}$.

Even though this expression \totalpp\  is clear from the equivalent
CHS geometry $\IR_{\phi}\times SU(2)_{n-2}$, it is instructive to
derive it from the viewpoint of the $N=2$ minimal model coupled to
the $N=2$ Liouville theory. We write the bosonic fields in the $N=2$
Liouville theory as $(Y,\phi)$. The matter partition function
\defpart\ is divided into the Liouville part and the other
contributions $Z_{\hat{c}=2}=Z_{L}Z_{others}$.  The former is simply
given by \eqn\liouvp{Z_{\phi}=\log\mu_{sl}\cdot
\f{1}{\s{\tau_2}|\eta(\tau)|^2}~.} By imposing that the R-charge is
integral, we have $m/N-Qp\in Z$, where $(l,m)$ is the primary
\dsursu\ of $N=2$ minimal model and $p$ is the $Y-$momentum. This
projection is equivalent to the $\IZ_n$ orbifold in \cosetdefg.
 Then we find
\eqn\otherp{\eqalign{Z^{(l)}_{others}&=\sum_{l=0}^{N-2}\sum_{m\in
Z_{2N}}\f{\theta_3(z|\tau)}{\eta(\tau)}
ch^{NS}_{lm}(\tau,z)\f{\Theta_{m,N}(\tau,-2z/N)+
\Theta_{m+N,N}(\tau,-2z/N)}{\eta(\tau)}\cr
&=\sum_{l=0}^{N-2}\f{\theta_3(z|\tau)}{\eta(\tau)^2}
\chi^{(l)}(\tau,0)~,}} where we adopt the convention and applied an
important identity in \ESP.  Each three factors in the second
expression in \otherp\ are the contributions from the fermions in
the $N=2$ Liouville, the $N=2$ minimal model and the $Y$ boson,
respectively.

\listrefs

\bye